\documentclass[a4paper,11pt]{article}

\pdfoutput=1

\usepackage{jcappub}

\usepackage[english]{babel}
\usepackage{units}
\usepackage[separate-uncertainty = true]{siunitx}
\usepackage{graphicx}
\usepackage{amsmath}
\usepackage{amssymb}
\usepackage{subfigure}
\usepackage{lineno}
\usepackage{multirow}
\usepackage{tikz}
\usepackage{pgf}
\usepackage{xspace}

\definecolor{dark-gray}{gray}{0.3}

\title{Neutrino vertex reconstruction with in-ice radio detectors using surface reflections and implications for the neutrino energy resolution}

\author[a]{A.~Anker}
\author[a]{S.~W. Barwick}
\author[b]{H. Bernhoff}
\author[c,d]{D.~Z. Besson}
\author[e]{N. Bingefors}
\author[f,g]{D. Garc\'ia-Fern\'andez}
\author[a]{G. Gaswint}
\author[a, 1]{C. Glaser\note{Corresponding author}}
\author[e]{A. Hallgren}
\author[h]{J. C. Hanson}
\author[i]{S.~R. Klein}
\author[j]{S.~A. Kleinfelder}
\author[a,g]{R. Lahmann}
\author[c]{U. Latif}
\author[k]{J. Nam}
\author[c,d]{A. Novikov}
\author[f,g]{A. Nelles}
\author[a]{M. P. Paul}
\author[a]{C. Persichilli}
\author[f,g]{I. Plaisier}
\author[i]{T. Prakash}
\author[a]{S. R. Shively}
\author[a,l]{J. Tatar}
\author[e]{E. Unger}
\author[k]{S.-H. Wang}
\author[f, g]{C. Welling}
\author[m]{S. Zierke}

\affiliation[a]{Department of Physics and Astronomy, University of California, Irvine, CA 92697, USA}
\affiliation[b]{Uppsala University Department of Engineering Sciences, Division of Electricity, Uppsala, SE-752 37 Sweden}
\affiliation[c]{Department of Physics and Astronomy, University of Kansas, Lawrence, KS 66045, USA}
\affiliation[d]{National Research Nuclear University MEPhI (Moscow Engineering Physics Institute), Moscow 115409, Russia}
\affiliation[e]{Uppsala University Department of Physics and Astronomy, Uppsala, SE-752
37, Sweden}
\affiliation[f]{DESY, 15738 Zeuthen, Germany}
\affiliation[g]{ECAP, Friedrich-Alexander-Universit\"at Erlangen-N\"urnberg, 91058 Erlangen, Germany}
\affiliation[h]{Whittier College Department of Physics, Whittier, CA 90602, USA}
\affiliation[i]{Lawrence Berkeley National Laboratory, Berkeley, CA 94720, USA}
\affiliation[j]{Department of Electrical Engineering and Computer Science, University of California, Irvine, CA 92697, USA}
\affiliation[k]{Department of Physics and Leung Center for Cosmology and Particle Astrophysics, National Taiwan University, Taipei 10617, Taiwan}
\affiliation[l]{Research Cyberinfrastructure Center, University of California, Irvine, CA 92697 USA}
\affiliation[m]{III. Physikalisches Institut, RWTH Aachen University, Aachen, Germany}

\emailAdd{christian.glaser@uci.edu}

\abstract{
    Ultra high energy neutrinos ($E_\nu > $\SI{e16.5}{eV}) are efficiently measured via radio signals following a neutrino interaction in ice. An antenna placed $\mathcal{O}$(\SI{15}{m}) below the ice surface will measure two signals for the vast majority of events (90\% at $E_\nu$=\SI{e18}{eV)}: a direct pulse and a second delayed pulse from a reflection off the ice surface. This allows for a unique identification of neutrinos against backgrounds arriving from above. Furthermore, the time delay between the direct and reflected signal (D'n'R) correlates with the distance to the neutrino interaction vertex, a crucial quantity to determine the neutrino energy. In a simulation study, we derive the relation between time delay and distance and study the corresponding experimental uncertainties in estimating neutrino energies. We find that the resulting contribution to the energy resolution is well below the natural limit set by the unknown inelasticity in the initial neutrino interaction. 
    We present an in-situ measurement that proves the experimental feasibility of this technique. Continuous monitoring of the local snow accumulation in the vicinity of the transmit and receive antennas using this technique provide a precision of $\mathcal{O}$(\SI{1}{mm}) in surface elevation, which is much better than that needed to apply the D'n'R technique to neutrinos.} 
   
\begin{document}
\maketitle
\flushbottom

\section{Introduction}

High-energy neutrino astronomy opens a new window to the universe and its most violent processes \cite{Astro2020NeutrinoAstronomy}. Neutrinos are undeflected in their journey through the universe and point back to their sources. In particular, a multi-messenger observation, including neutrinos together with electromagnetic measurements ranging from radio to optical to gamma rays allows to pinpoint and better understand the sources. The multi-messenger era was initiated when the IceCube detector at the South Pole observed a \SI{3e14}{eV} neutrino in coincidence with a flaring blazar observed with gamma-ray telescopes \citep{IceCubeBlazer2018}. Additional neutrino detectors with a similar sensitivity are currently being constructed in the Mediterranean sea \cite{KM3Net2016} and Lake Baikal \cite{GVD2019} to observe the Northern hemisphere. 

To extend the detectable neutrino energy range beyond \SI{e16}{eV}, which will potentially link neutrinos with gravitational wave observations \cite{Kimura2017, Fang2017a}, one needs a different detection technology. The radio technique allows to cost-efficiently instrument large volumes \citep{IceCubePRL2016} to cover neutrino energies above \SI{e16.5}{eV} due to the large attenuation length of radio signals in ice of $\mathcal{O}$(\SI{1}{km}). A neutrino interacting in the ice produces a particle cascade that generates a short radio pulse over the frequency range \SI{50}{MHz} to \SI{1}{GHz} via the Askaryan effect \cite{Askaryan1962}. 

This promising technique was successfully explored in two pilot arrays ARA \cite{ARA} at the South Pole and ARIANNA \cite{ARIANNACospar2018} on the Ross ice shelf and at the South Pole. The requisite hardware and technology has matured in these pilot arrays and the construction of a large-scale detector with enough sensitivity to detect statistically significant numbers of ultra-high energy neutrinos is foreseen in the near future. 

With the transition from a pilot to production phase, the focus shifts from building a working detector with large sensitivity to determining the relevant neutrino properties, specifically direction and energy, from the observed few-nanosecond duration radio flashes. As the detectors are optimized for maximum sensitivity, most neutrinos will be observed in only a few antennas of a single detector station, which makes reconstruction of neutrino properties challenging. In this article, we explore in depth how the distance to the neutrino interaction vertex can be determined and then translated into a neutrino energy estimate. 

The distance to the neutrino vertex can be measured precisely via the D'n'R (direct and reflected) technique \cite{Allison2019,KelleyARENA2018}. An antenna placed $\mathcal{O}$(\SI{15}{m}) below the ice surface will measure two pulses for the vast majority of detected neutrinos \cite{NuRadioMC2019}, one direct signal and a slightly delayed signal that is reflected off the ice surface. The time delay between the two pulses is a proxy for the vertex distance. 
The geometries are such that most reflections are totally internally reflected (TIR) meaning that both pulses are comparable in amplitude. 
For simplicity we use the term 'vertex position' to refer to the point of emission although they are not exactly the same because of the extent of the particle shower (see also discussion in Sec.~\ref{sec:theory}).

We first perform a Monte-Carlo study using NuRadioMC \cite{NuRadioMC2019} in which we determine the relation between time delay and distance, and simulate the resolution on the vertex distance and the corresponding contribution to the neutrino energy uncertainty. Then, we present an in-situ measurement performed with the ARIANNA detector on the Ross ice shelf to demonstrate the experimental feasibility of this technique. In the last section, we demonstrate how this technique can be used to continuously monitor snow accumulation; we also present a several-month snow accumulation measurement.

\section{Energy reconstruction}
\label{sec:y}
In this section, the steps necessary to reconstruct the neutrino energy from a radio detector are briefly summarized and the performance required of the D'n'R technique to determine the vertex distance are established.

The relation between the neutrino energy ($E_\nu$) and the radio signal amplitude observed at the detector ($\varepsilon$), for a neutrino interaction at a range R from the detector is summarized in the following equation
\begin{equation}
    \varepsilon(f) = \varepsilon_0(f, E_\nu, y, ...) \times \frac{e^{-R/L(f)}}{R} \times \exp\left[\frac{-(\theta - \theta_C)^2}{2 \sigma_\theta(f)^2}\right] \, ,
    \label{eq:y}
\end{equation}
where $\varepsilon$ describes the observed frequency spectrum of the Askaryan signal, $\varepsilon_0$ is the Askaryan signal as a function of neutrino energy $E_\nu$, inelasticity $y$ and other properties of the neutrino interaction. The parameter $L(f)$ is the frequency dependent attenuation length and $R$ the distance. The angles $\theta$ and $\theta_C$ are the viewing angle and the Cherenkov angle respectively, and $\sigma_\theta$ is the width of the Cherenkov cone. Please refer to \cite{GlaserICRC2019} for a more detailed discussion of each of the terms.

The second and third terms on the right-hand side of this equation depend on the measurable quantities \emph{vertex distance $R$} and \emph{viewing angle $\theta$}. The first term represents the fraction of the primary neutrino energy manifest as radio emission. It depends on stochastic processes in the neutrino interaction and imposes an irreducible energy uncertainty. Thus, it sets the scale for the experimental precision required for the second and third terms in Eq.~\eqref{eq:y}.

The radio signal amplitude at the source scales linearly with the energy of the particle shower ($E_\mathrm{sh}$) generated following a neutrino interaction. 
Energy is transferred into the shower stochastically and depends on the specific type of interaction. For charged-current electron neutrino interactions, an electromagnetic shower is induced by an electron generated in the neutrino interaction, and a hadronic shower results from the interaction of the neutrino with the nucleus. For all other types of interactions, only a hadronic shower is created (we neglect decays of tau leptons and catastrophic dE/dx from muons for simplicity). We note that a hadronic shower will eventually transfer most of its energy into electromagnetic sub-showers which are responsible for the radio emission. Thus, with the term 'hadronic showers' we refer to a particle shower with initial hadronic interaction that will transfer most of its energy into electromagnetic sub-showers.   

The shower energy can be related to the neutrino energy via
\begin{align}
E_\mathrm{sh} =
\begin{cases}
y \, E_\nu \, &\text{for hadronic showers} \\
(1- y) \, E_\nu \, &\text{for electromagnetic showers}
\end{cases}
\end{align}
We note that only the shower energy that ends up in electromagnetic sub-showers is relevant to the radio emission. For hadronic showers only 90\% to 95\% of the energy ends up in electromagnetic cascades \cite{Alvarez-Muniz1998} which is precisely modelled in the Askaryan emission codes. 

The distribution of inelasticity $y$ (see e.g. \cite{Gandhi1996} or \cite{Connolly2011}) cannot be used directly as a proxy for the scatter in reconstructed neutrino energy because of the detector acceptance. An interaction with an inelasticity value that leads to a small shower energy is less likely to be detected than those showers for which a significant part of the neutrino energy is transferred to the shower. 

To study this effect under realistic conditions, we performed a full Monte Carlo simulation using NuRadioMC \cite{NuRadioMC2019}, i.e., we simulate the initial neutrino interaction, followed by radio signal generation and propagation to a detailed detector simulation. We simulate an initial neutrino energy spectrum following a $E^{-2.2}$ power law, corresponding to an extrapolation of the astrophysical neutrino flux measured by IceCube \cite{Haack2017}, superimposed upon a cosmogenic neutrino spectrum, i.e., neutrinos generated via cosmic-ray interactions with the cosmic microwave background, for a 10\% proton fraction and for a standard choice of source evolution \cite{Vliet2019} (see also \cite{ARIANNACospar2018} for a discussion of the models). 

\begin{figure}[t]
    \centering
    \includegraphics[width=0.49\textwidth]{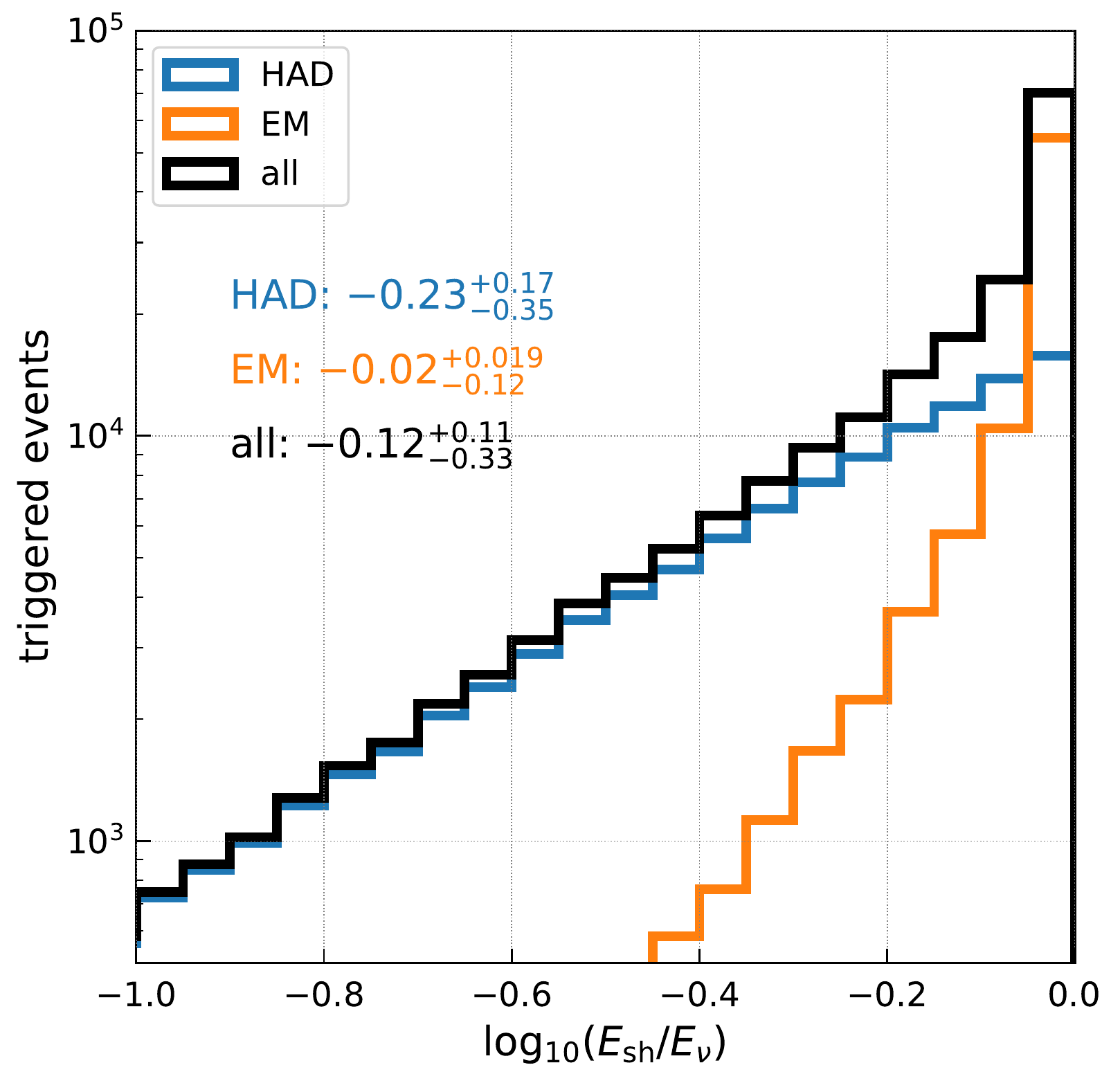}
    \includegraphics[width=0.49\textwidth]{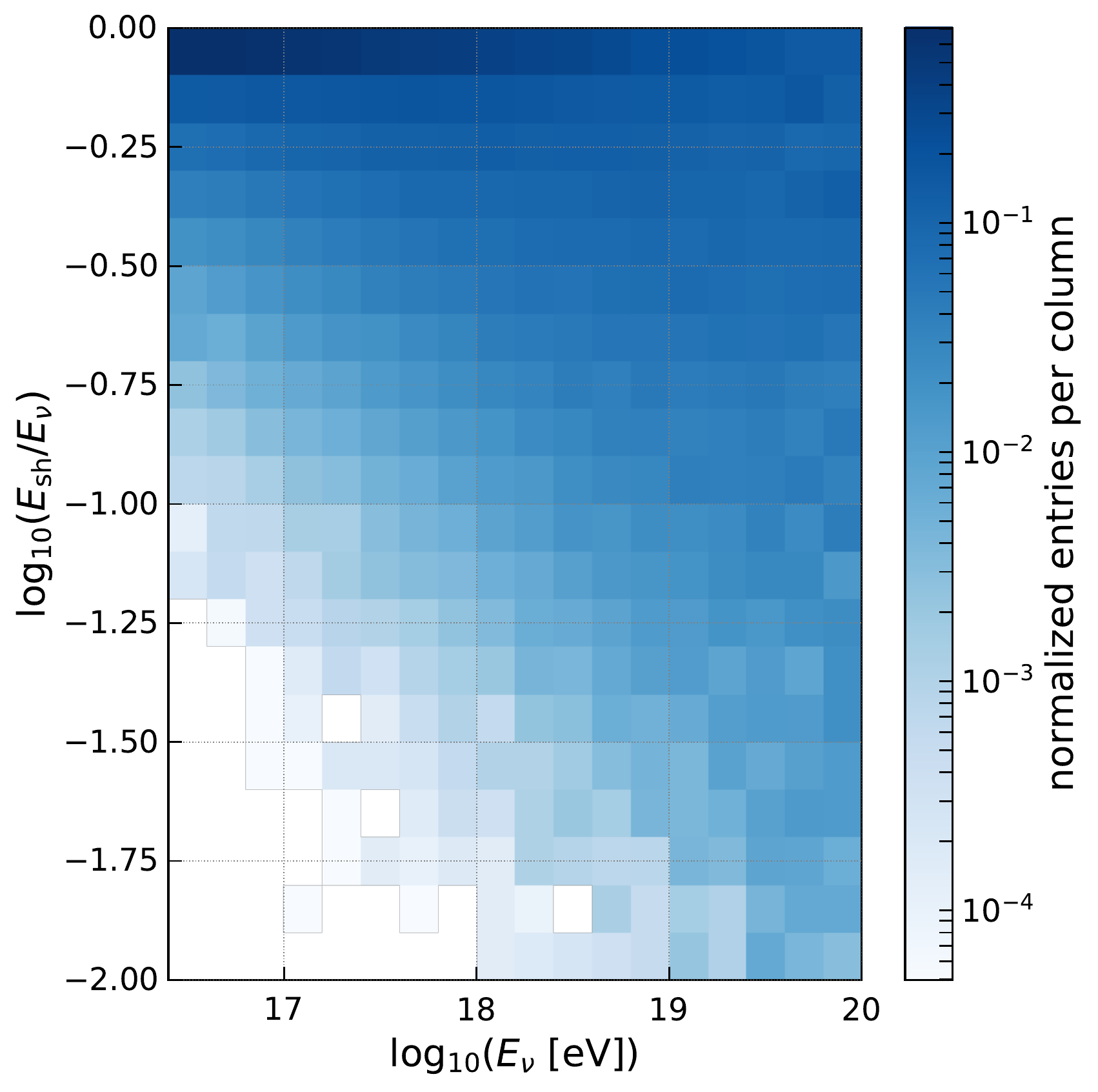}
    \caption{(left) Distribution of the ratio between shower energy and neutrino energy for triggered events. The solid curve is for an initial neutrino energy spectrum obtained by summing astrophysical plus cosmogenic signal components (see text for details). (right) Ratio between shower energy and neutrino energy as a function of neutrino energy for all triggered events.}
    \label{fig:Esh}
\end{figure}

In Fig.~\ref{fig:Esh} left, we present the ratio of shower energy to initial neutrino energy for all triggered events, and also separately for hadronic and electromagnetic showers. The distribution is strongly biased towards high transferred energy such that the shower energy can be close to the neutrino energy with the bias most pronounced for electromagnetic showers. The distribution is also energy dependent (see Fig.~\ref{fig:Esh} right), broadening with increasing neutrino energy because already a small energy transfer results in sufficiently energetic showers to trigger the detector.
The distribution in Fig.~\ref{fig:Esh} is clearly non-Gaussian; we choose to describe it via the median value and the 68\% quantiles which we calculate for the astrophysical + cosmogenic spectrum and discrete neutrino energies:
\begin{align}
\log_{10}(E_\mathrm{sh}/E_\nu) = 
\begin{cases}
-0.12^{+0.11}_{-0.33} \, &\text{for astrophysical + cosmogenic spectrum} \\
-0.06^{+0.06}_{-0.22} \, &\text{at \SI{e17}{eV} neutrino energy} \\
-0.25^{+0.18}_{-0.34} \, &\text{at \SI{e18}{eV} neutrino energy}\\
-0.33^{+0.26}_{-0.49} \, &\text{at \SI{e19}{eV} neutrino energy}
\end{cases}
\end{align}
We estimate the resulting uncertainty on the neutrino energy to be about $0.3$ in the logarithm of $\log_{10}(E_\mathrm{sh}/E_\nu)$, corresponding to a factor of $2$ on a linear scale.
This imposes a natural limit on the maximum experimentally achievable energy resolution for high-energy neutrino detection, and sets the scale for the optimal experimental precision: The uncertainty of the vertex distance and viewing angle should be small enough so as not to significantly increase the energy uncertainty beyond this inelasticity limit. 

We note that for a subset of detected events the neutrino energy might be determined more precisely. At high neutrino energies ($E_\nu > \SI{e18}{eV}$) electromagnetic and hadronic showers might be differentiated: Electromagnetic showers are elongated by the LPM effect \cite{Gerhardt:2010bj}, resulting in a reduced Cherenkov cone width that can be measured with an array of antennas with enough spatial separation. Furthermore, for $\nu_e$ charge-current interactions, both the electromagnetic and hadronic showers might be detected either if both showers are sufficiently spatially displaced \cite{Aartsen2019} or by measuring the distinct frequency spectrum with broadband antennas. In this case the inelasticity uncertainty can be removed completely as the sum of hadronic and electromagnetic shower energy gives the neutrino energy. 
As these signatures will be measurable only for a small fraction of neutrino events, we ignore them in the following discussion and use the irreducible limit on the neutrino energy resolution of $0.3$ in $\log_{10}(E_\mathrm{sh}/E_\nu)$ as a lower bound on the achievable precision.

\section{Determination of neutrino vertex position}
\label{sec:theory}
In this section, we study how to determine the distance to the neutrino vertex by measuring the time delay between the direct and reflected signal path.
At the end of this section, we estimate the expected vertex resolution and how this propagates into the neutrino energy resolution. 

\begin{figure}[t]
    \centering
    \includegraphics[width=0.6\textwidth]{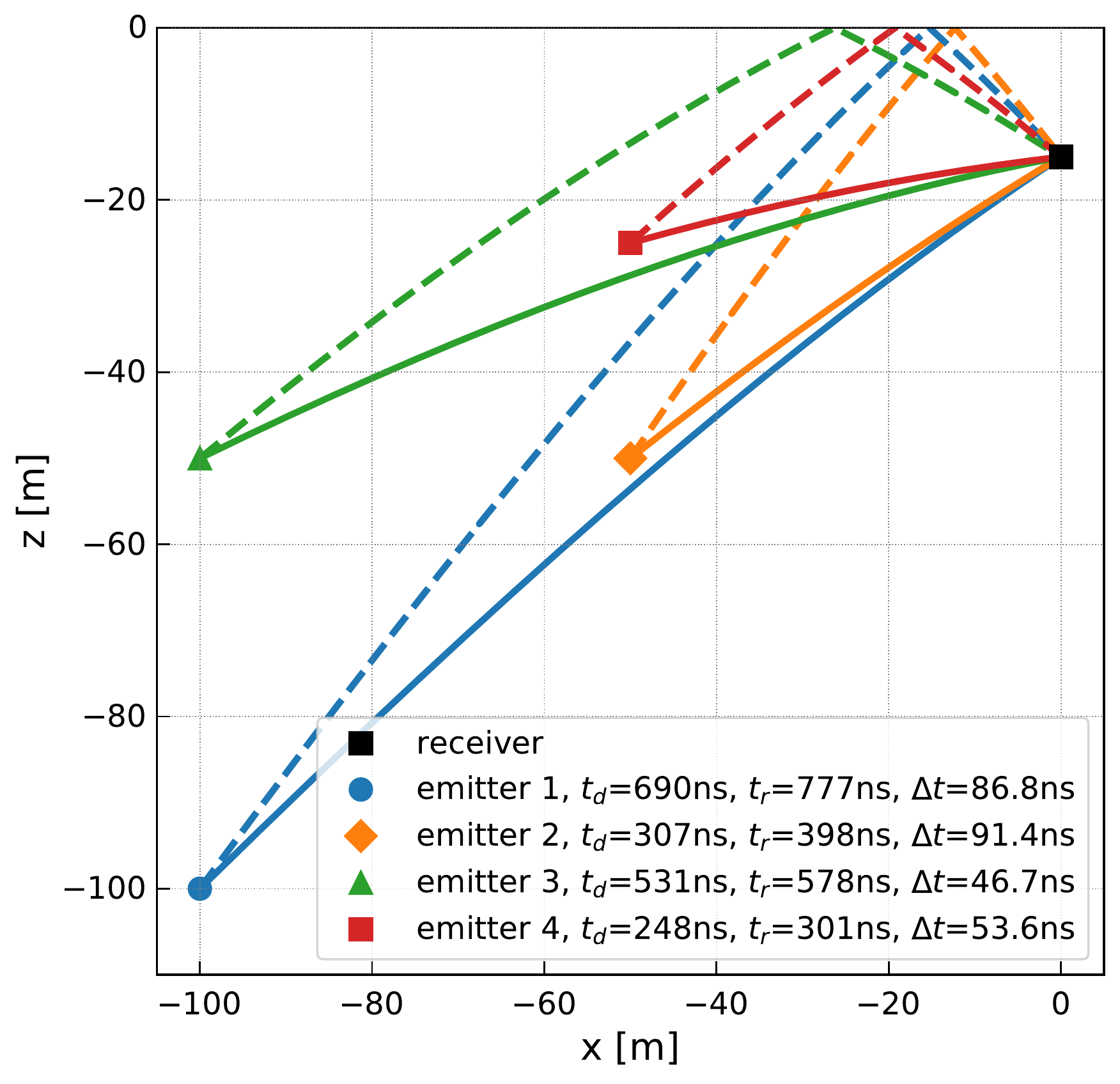}
    \caption{Illustration of typical signal paths. The solid curves show direct rays; the dashed curves show rays that are reflected off the snow-air surface. The legend specifies the propagation time of direct rays $t_d$ and reflected rays $t_r$ as well as their time difference  $\Delta t$. Here, we show relatively close emitter positions for better readability. Typically, neutrino vertices will be further away (cf. Fig.~\ref{fig:MCdist}).}
    \label{fig:demo}
\end{figure}

We illustrate a few typical signal paths in Fig.~\ref{fig:demo}. 
The time delay $\Delta t$ between the direct and reflected ray depends on the distance between receiver and emitter, their depths, and the incoming signal direction. In all calculations, we take into account the change in the speed-of-light at different depth resulting from the changing index-of-refraction profile. Similarly, all quoted distances are determined along the curved signal paths dictated by Fermat's least-time principle. 

With the additional information of the radio-frequency signal arrival direction, which is well-known experimentally (e.g. \cite{Allison2019,ARIANNACospar2018}), the neutrino vertex position can be determined by following the signal path, defined by the arrival direction, backwards. Hence, once the signal path is prescribed, the vertex distance prescribes vertex position (and vice versa). We note that the neutrino interaction vertex is not exactly the position where the Askaryan signal originates from because of the extent of the initiated particle shower. Most radio signal is emitted at the maximum of the particle shower which is $\mathcal{O}(\SI{10}{m})$ closer to the receiver than the interaction vertex. The exact displacement depends on energy and the degree of LPM elongation \cite{NuRadioMC2019}. For simplicity, however, we ignore this subtle difference and refer to the point of emission as the `vertex position'. 

In the following, we will first focus on a \SI{15}{m} deep receiver which is a good compromise between high efficiency to detect both D'n'R pulses and good time resolution on their separation in a waveform. Later, we extend this study to include a range of possible receiver depths.

We simulated 70 million vertex positions randomly placed within a cylinder of \SI{5}{km} radius and \SI{2.7}{km} depth (the thickness of the ice sheet at South Pole), and calculated the signal trajectories to a \SI{15}{m} deep receiver. We used the fast analytic ray tracer of NuRadioMC \cite{NuRadioMC2019} to calculate the signal trajectories and the propagation times using the \emph{SPICE 2015} parameterization of the index of refraction profile of South Pole from \cite{Barwick2018}. For a shallow receiver, most vertex positions have either no possible signal path to reach the receiver, or a direct and reflected signal path to the receiver. Geometries with a refracted and reflected path or two refracted signal paths will be ignored in the following. For a realistic simulation of expected neutrino signals, the fraction of such events is only 4\%. 

\begin{figure}[t]
    \centering
    \includegraphics[width=0.45\textwidth]{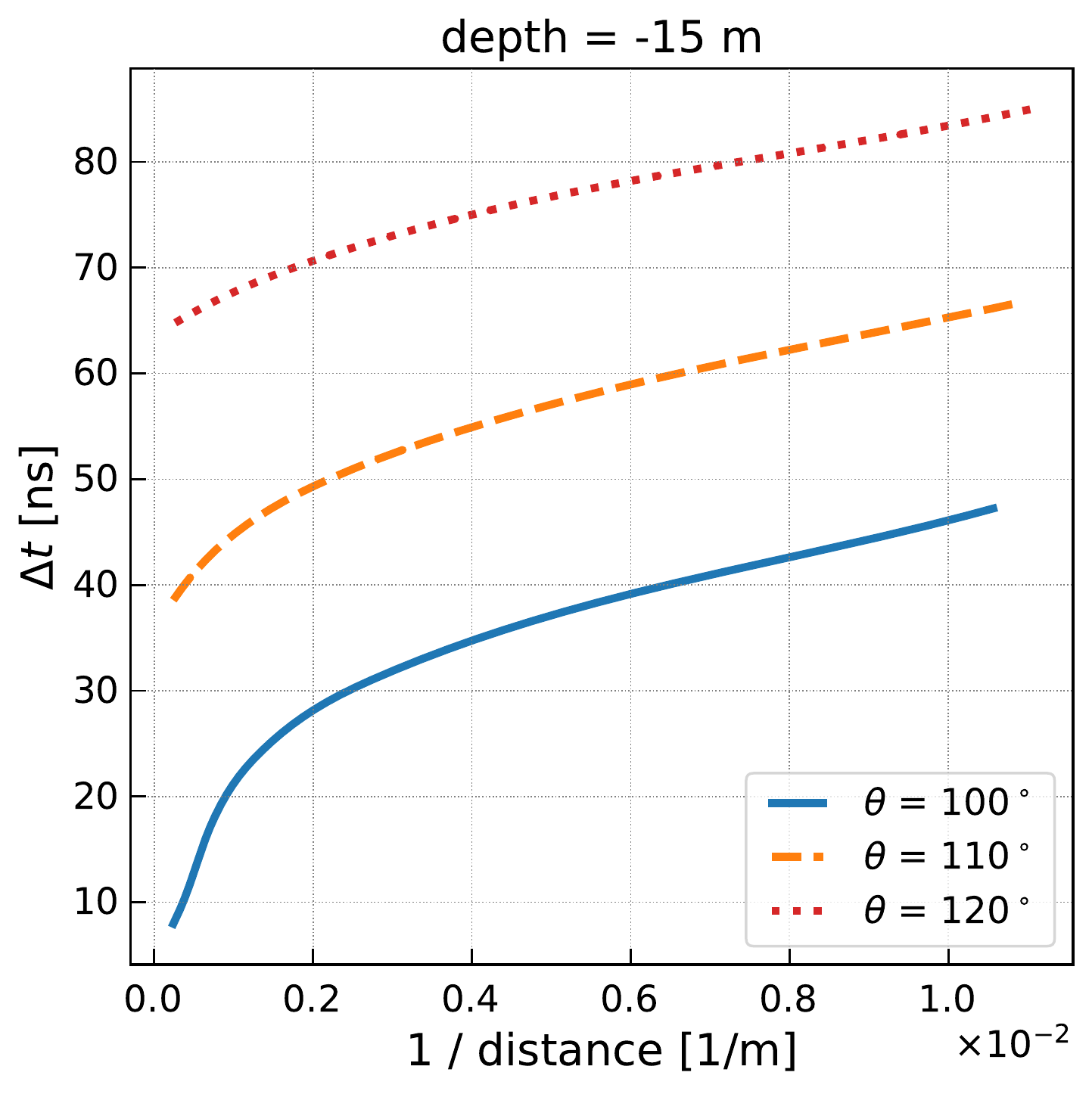}
    \includegraphics[width=0.54\textwidth]{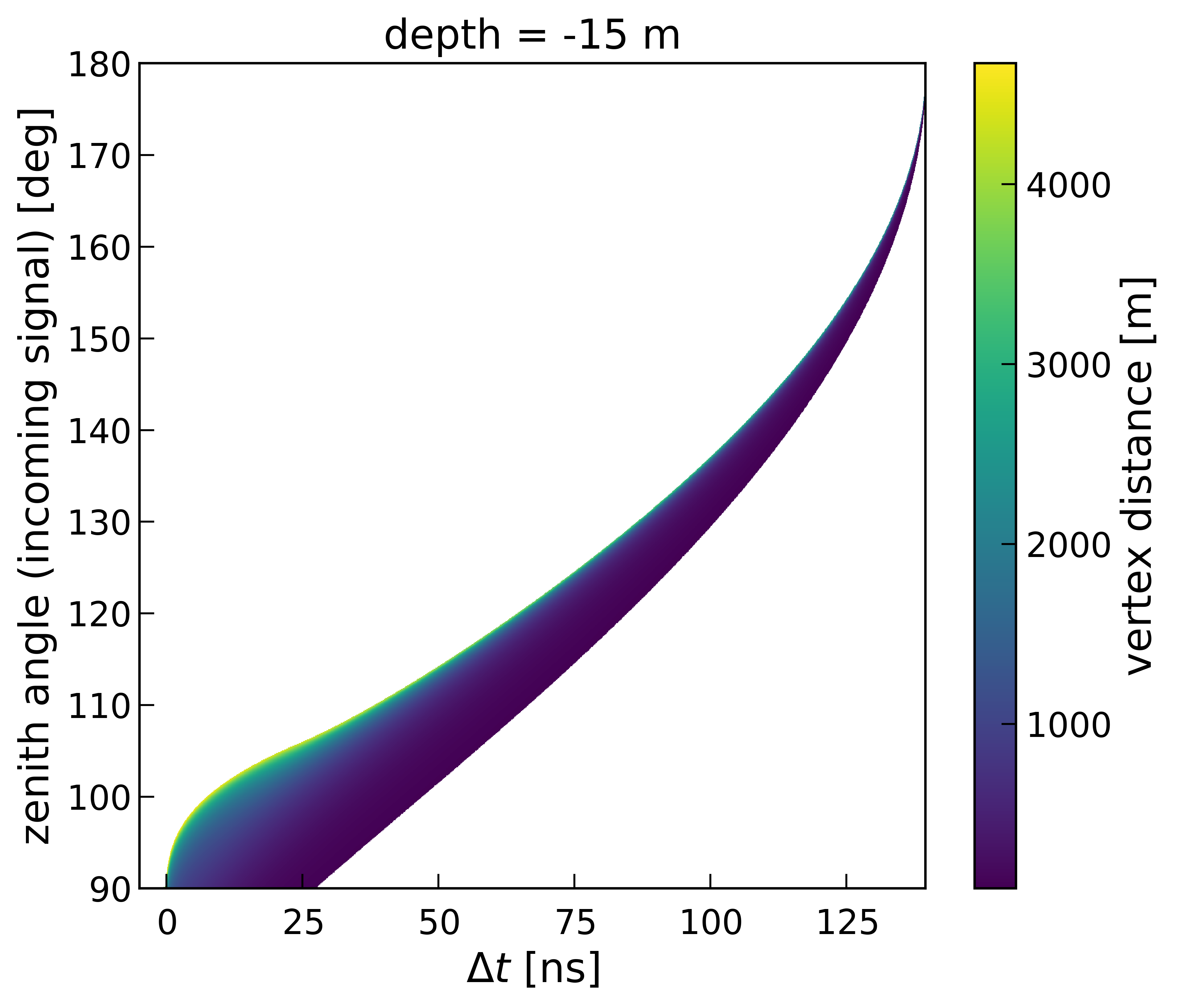}
    \caption{(left) Time delay between direct and reflected signal as a function of the inverse distance to the neutrino vertex, for three incoming signal directions. (right) Distance to the neutrino vertex (color coded) as a function of the incoming signal direction and time delay.}
    \label{fig:parameterization}
\end{figure}

The time delay as a function of distance for three incoming directions is presented in the left panel of Fig.~\ref{fig:parameterization}. The incoming direction is defined in terms of the zenith angle $\theta$. Because of radial symmetry, the azimuth direction is irrelevant here.  The time delay inversely correlates with the distance to vertex, as well as the signal elevation angle. We bin all simulations in \SI{0.1}{\degree} zenith angle and \SI{0.1}{ns} $\Delta t$ steps and illustrate the dependence on the vertex distance in the right panel of Fig.~\ref{fig:parameterization}. The bin widths are chosen to be smaller than the typical experimental uncertainties such that the binning does not limit the vertex resolution. This 2D profile serves as a lookup table to quickly translate a measured time delay and signal direction into vertex distance. 

\subsection{Vertex distance resolution}
The resolution on vertex distance depends not only on the $\Delta t$ and zenith angle $\theta$ resolution but also on the vertex distance and incoming signal direction itself. This is because the slope of $\Delta t$ vs. distance $R$ is smaller for distant vertices and vertically arriving signals. Thus, nearby vertices will have better resolution than distant vertices. Similarly, incoming directions from close to the horizon have better distance resolution than signals arriving from straight down. Therefore, we fold in the expected vertex distribution of neutrinos and corresponding incoming signal directions. The distribution depends on the neutrino energy, with larger neutrino energies having more distant vertices on average. Hence, we study the resolution for a fixed neutrino energies of \SI{e17}{eV} and \SI{e18}{eV}, corresponding to the peak sensitivity of an Askaryan detector. 

\begin{figure}[t]
    \centering
    \includegraphics[height=0.38\textwidth]{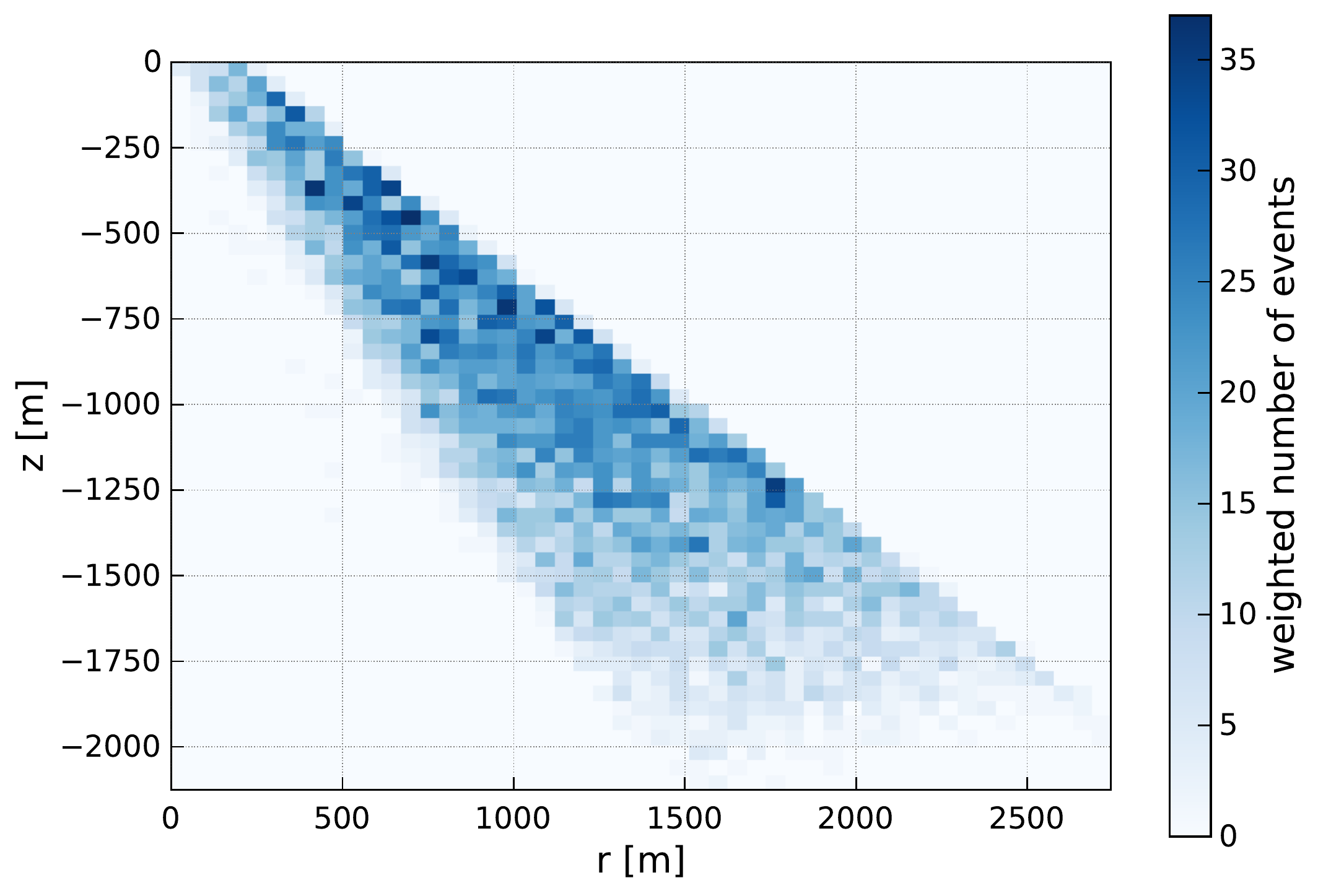}
    \includegraphics[height=0.38\textwidth]{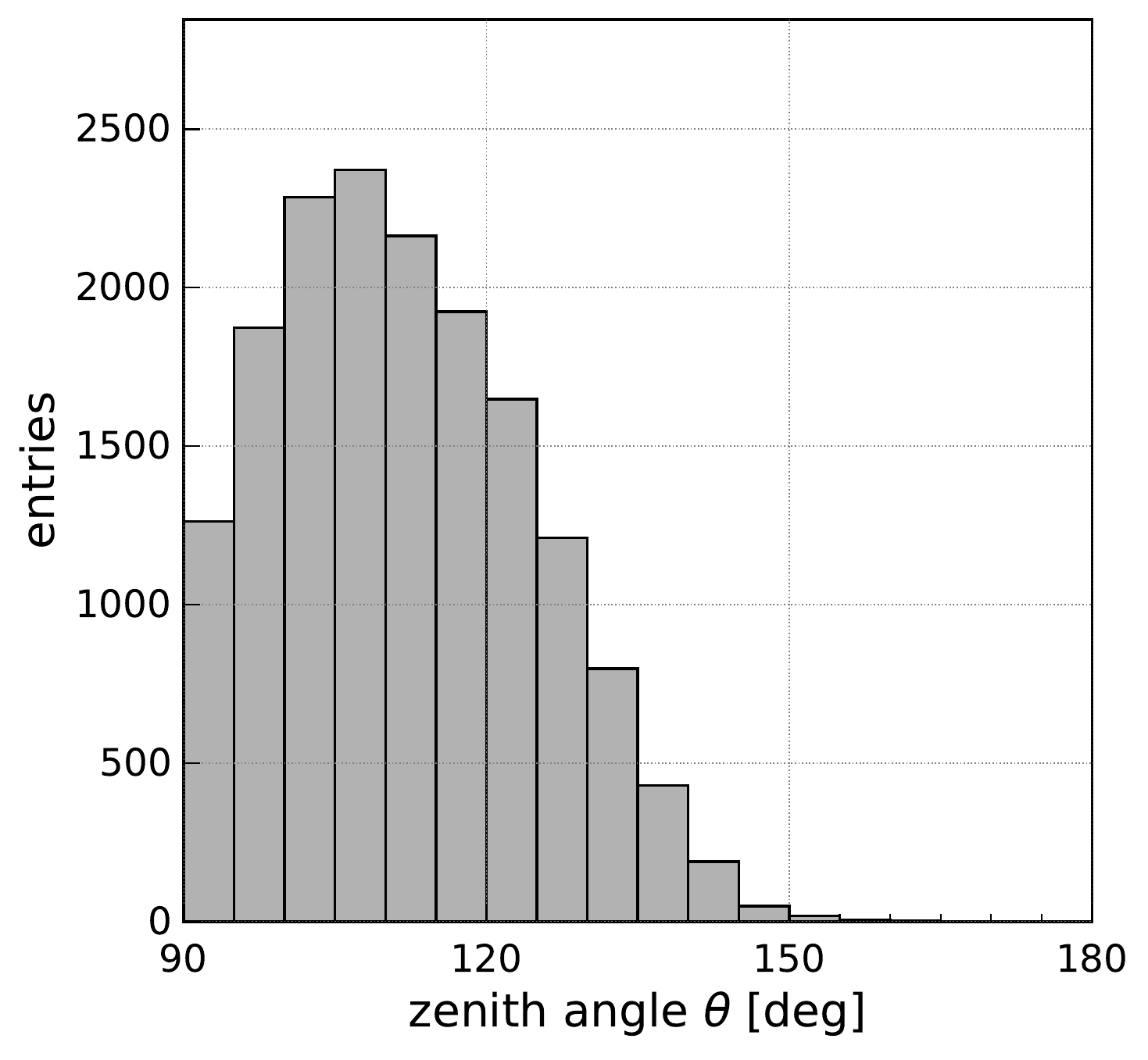}
    \caption{Vertex distribution (left) and zenith angle distribution of the incoming signal direction (of the direct signal path) (right) from a full MC simulation using NuRadioMC for a fixed neutrino energy of \SI{e18}{eV} and a receiver depth of \SI{15}{m}. A zenith angle of \SI{90}{\degree} points to the horizon and \SI{180}{\degree} points straight down.}
    \label{fig:MCdist}
\end{figure}

Fig.~\ref{fig:MCdist} presents the expected distribution of vertex positions and incoming signal directions for high-energy neutrinos to illustrate the relevant parameter space. We obtained these distributions from NuRadioMC, employing a detailed calculation of the Askaryan signal using the \emph{Alvarez2009} model \cite{Alvarez2009, Alvarez2012}, signal propagation including ice attenuation effects, and a full detector simulation. (This simulation corresponds to Example 2 of Ref.~\cite{NuRadioMC2019}.) The distribution of incoming signal directions (for the direct signal path) is favorable for obtaining good distance resolution, as most signals arrive from close to the horizon, where the $\Delta t$ vs. distance dependence is large. The shape of the distribution a result of the Earth being opaque at these neutrino energies and because the emission is mostly concentrated around the Cherenkov cone with an opening angle of \SI{54}{\degree}. Thus, for the corner case of a neutrino coming from the horizon, the signal can be emitted upwards with a maximum zenith angle of \SI{90}{\degree} + \SI{54}{\degree} = \SI{144}{\degree}. Furthermore, the signal trajectories are bent downwards due to the change in the index-of-refraction profile resulting in more horizontal incoming directions. 

We determine the vertex distance resolution as follows: For each event of the NuRadioMC simulation that triggered the detector, the true zenith angle and $\Delta t$ is smeared 200 times according to a Gaussian-distributed uncertainty in  $\sigma_\theta$ and $\sigma_{\Delta t}$. The histogram of Fig.~\ref{fig:parameterization} (right) is used to look up the corresponding vertex distance, which is then compared to the true vertex distance. 

In Fig.~\ref{fig:resolution} (left), we present the vertex-distance resolution assuming a $\Delta t$ resolution of \SI{0.2}{ns} and a zenith angle resolution of \SI{0.2}{\degree}, corresponding to our estimate for the achievable experimental resolution of a future Askaryan detector \cite{ARIANNACospar2018}. As expected, the resolution is better for lower neutrino energies. For \SI{e17}{eV}, we find a 68\% quantile of 0.04 in $\log{10}(R_\mathrm{rec}/R_\mathrm{true})$, corresponding to a linear resolution of 10\%. For \SI{e18}{eV}, we find a 68\% quantile of 0.05 in $\log{10}(R_\mathrm{rec}/R_\mathrm{true})$, corresponding to a linear resolution of 12\%. 

\subsection{Energy resolution}
\begin{figure}[t]
    \centering
    \includegraphics[width=0.49\textwidth]{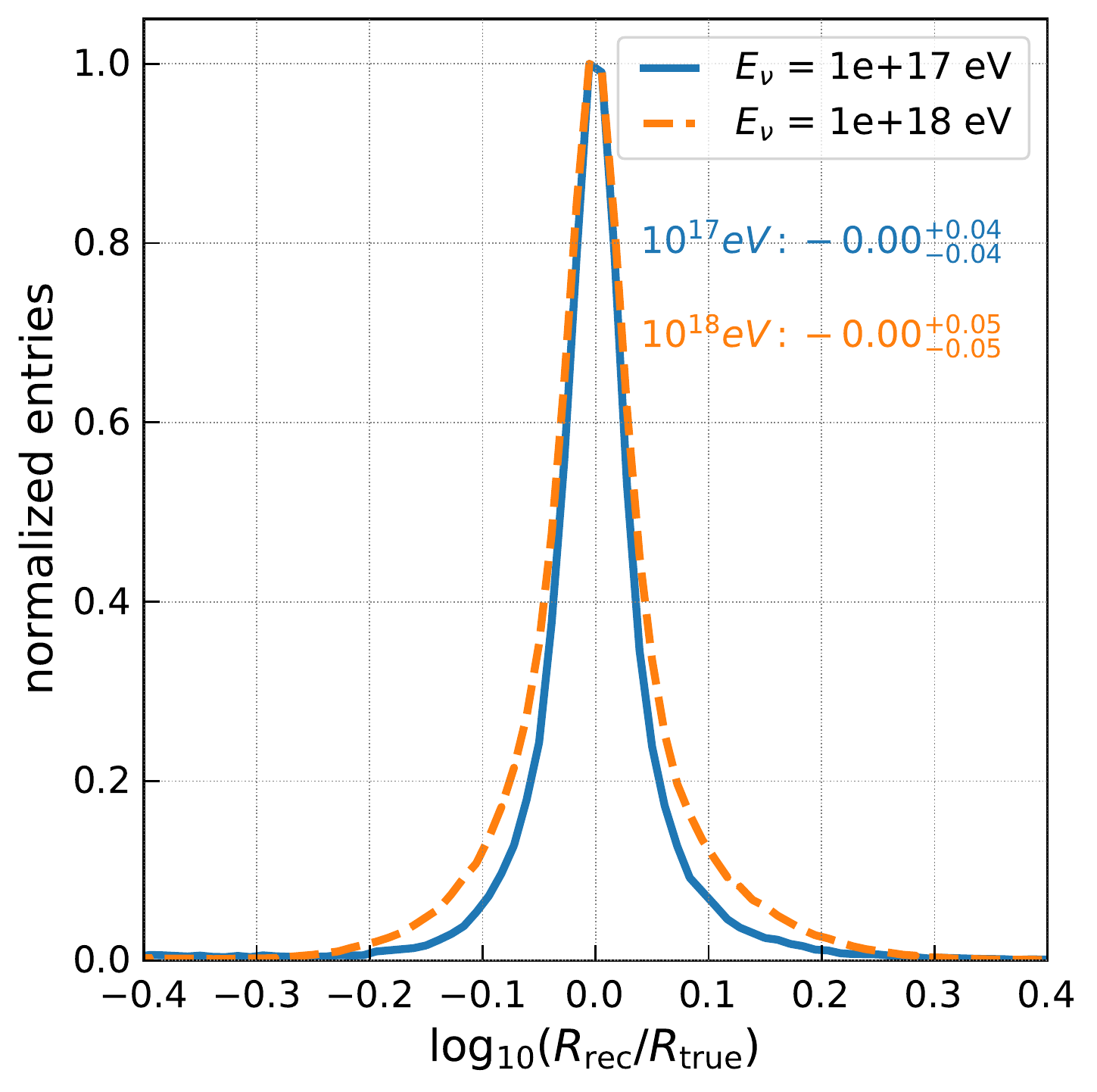}
    \includegraphics[width=0.49\textwidth]{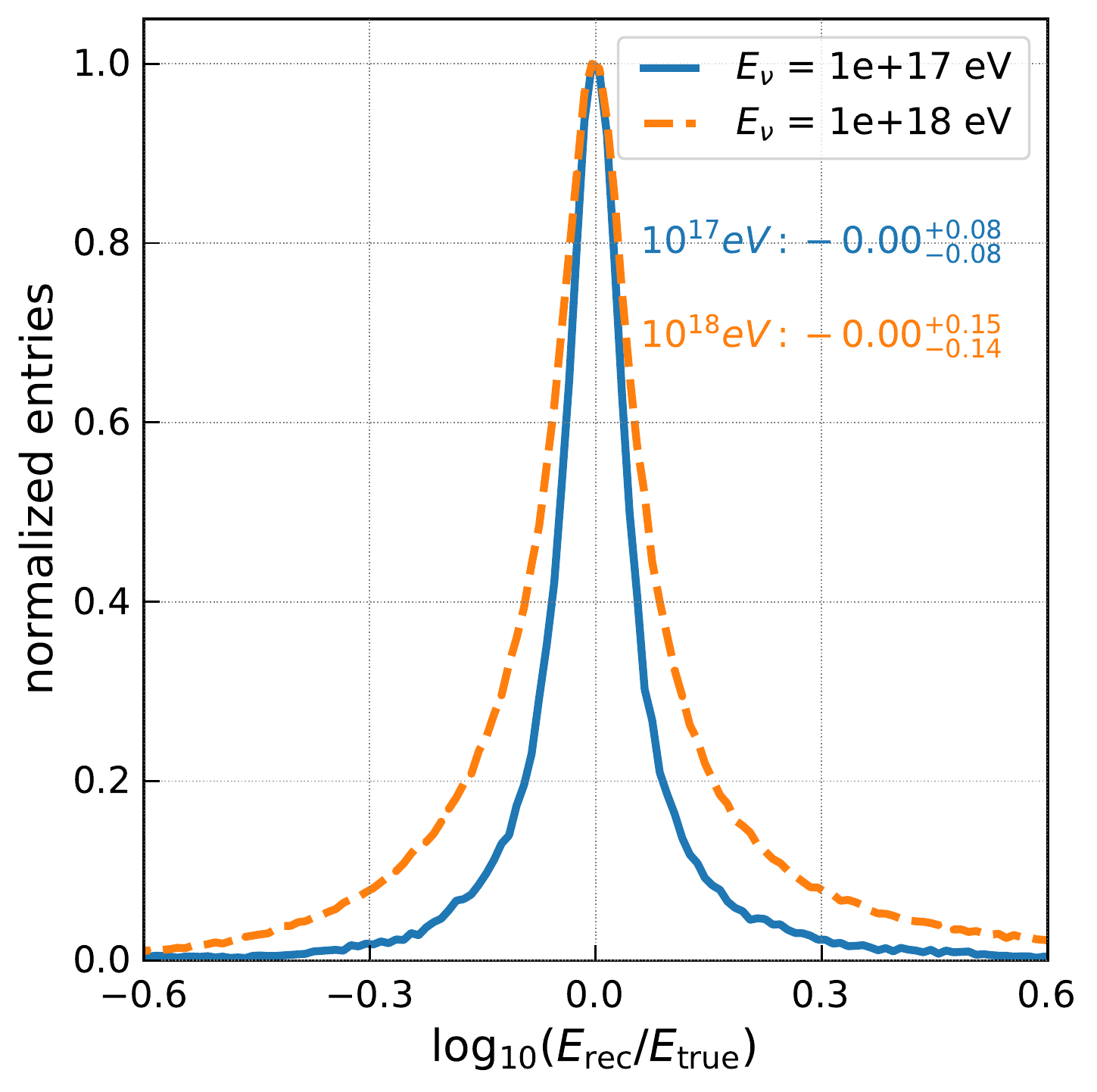}
    \caption{(left) Vertex distance resolution for a \SI{-15}{m} deep receiver and uncertainties of \SI{0.2}{ns} in the D'n'R time delay and \SI{0.2}{\degree} in the zenith direction. (right) Corresponding contribution to the energy resolution from uncertainties in the vertex distance.}
    \label{fig:resolution}
\end{figure}

For multi-messenger science, the relevant quantity of interest is not the vertex resolution but the neutrino energy resolution. Thus, for each vertex distance we calculate the 'shower energy' as 
\begin{equation}
 E_\mathrm{sh} \propto \frac{R}{\exp(-R/L_\mathrm{a})}\, ,
\end{equation}
where $R$ is the distance from the vertex to the antenna along the direct ray path and $L_\mathrm{a}$ is the attenuation length. This formula is essentially correcting a unit measured signal for attenuation. We use an attenuation length of \SI{1}{km} matching measurements from the South Pole \cite{Barwick2005, NuRadioMC2019}. The resulting energy resolution (from uncertainties of the vertex distance only) is presented in the right panel of Fig.~\ref{fig:resolution}. For neutrino energies of \SI{e17}{eV}, we find a resolution of $\log_{10}(E_\mathrm{rec}/E_\mathrm{true})$ of $\sim$0.08, corresponding to 20\% on a linear scale. For \SI{e18}{eV}, we find a resolution of +0.15 and -0.14 in $\log_{10}(E_\mathrm{rec}/E_\mathrm{true})$, corresponding to 38\% - 41\% on a linear scale. 
Thus, the energy uncertainty from the vertex distance is significantly smaller than the natural limit imposed by the inelasticity of the initial neutrino interaction of 0.3 in $\log_{10}(E_\mathrm{rec}/E_\mathrm{true})$ (cf. Sec.~\ref{sec:y}). 

\begin{figure}[t]
    \centering
    \includegraphics[width=0.49\textwidth]{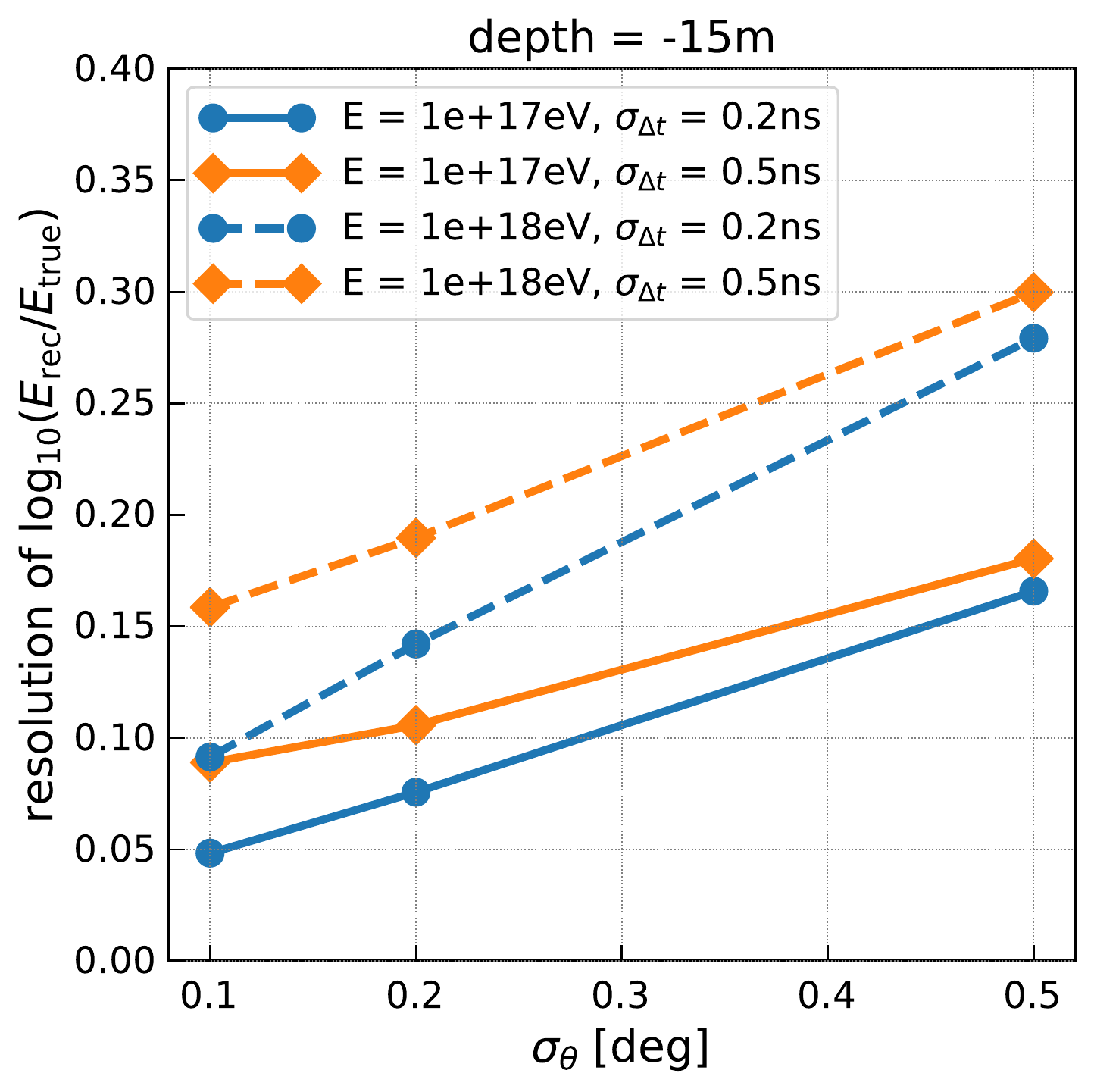}
    \includegraphics[width=0.49\textwidth]{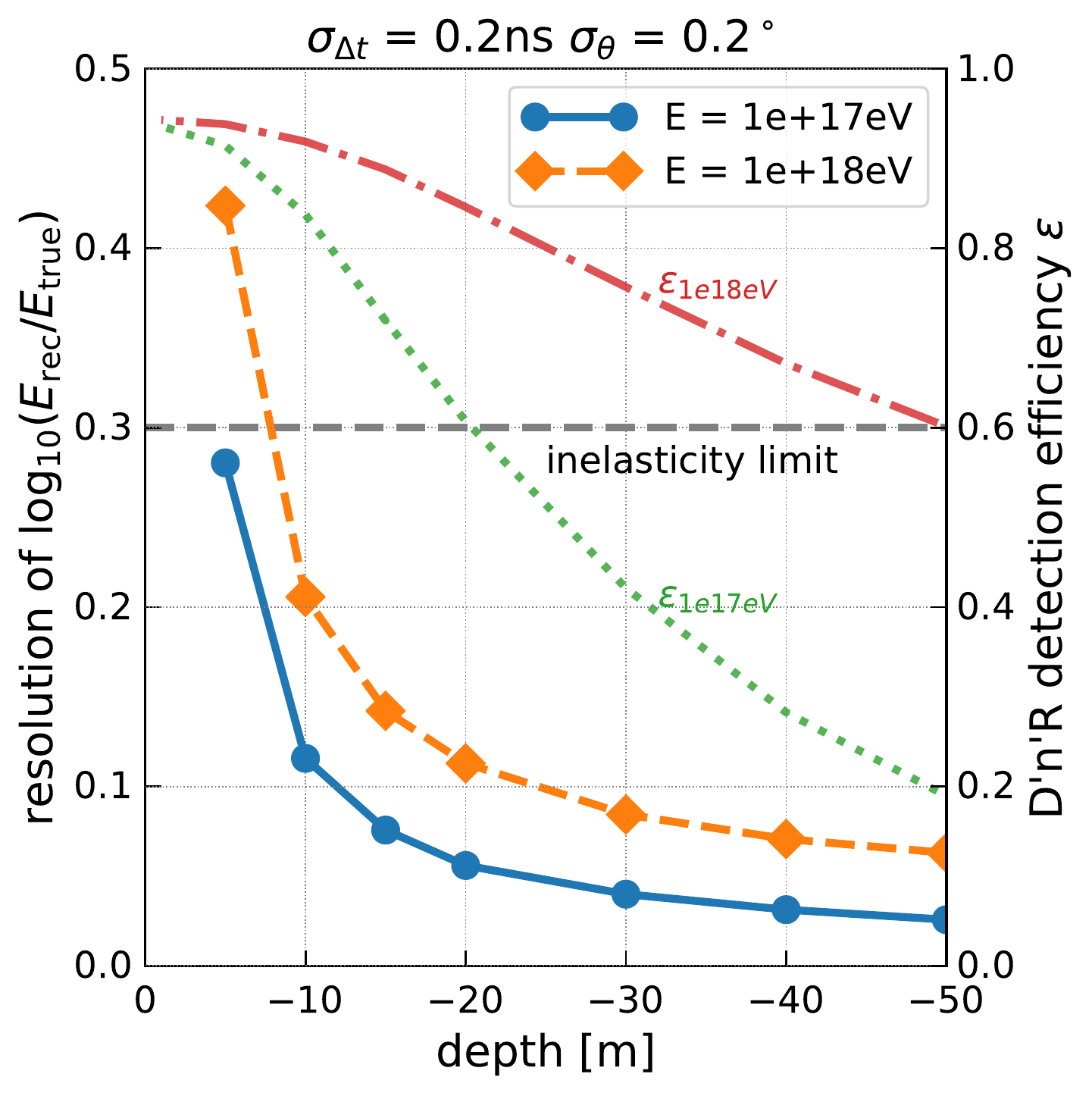}
    \caption{Dependence of energy resolution on $\Delta t$ and $\theta$ resolution (left) and dependence on depth (right). To interpret this figure, note that the efficiency to detect both D'n'R pulses decreases with depth. This is shown with the dotted (E = \SI{e17}{eV}) and dash-dotted (E = \SI{e18}{eV}) curves (see the right y-axis). Also shown is the limit on the energy resolution due to the unknown inelasticity, as a dashed horizontal line.}
    \label{fig:res_dep}
\end{figure}

In Fig.~\ref{fig:res_dep} (left), we present the dependence of the energy resolution on the uncertainty in $\Delta t$ and $\sigma_\theta$. Even for larger uncertainties of \SI{0.5}{\degree} and \SI{0.5}{ns}, the resulting energy resolution is still well below the inelasticity limit for \SI{e17}{eV} neutrino energies.

In Fig.~\ref{fig:res_dep} (right), we show the energy resolution as a function of receiver depth. Over the first $\sim$\SI{10}{m} the resolution improves dramatically. At greater depths, the resolution is continuously improving, but the relative improvement diminishes. Already at \SI{10}{m} depth, the contribution to the energy resolution from the vertex distance uncertainty is well below the natural limit of the energy resolution from the unknown inelasticity. 

To draw conclusions for an optimal detector layout from the depth dependence, the efficiency to detect both D'n'R pulses must also be considered. This was already studied in \cite{NuRadioMC2019} and is also shown in Fig.~\ref{fig:res_dep} (right) for the case of a $3 \, V_\mathrm{RMS}$ trigger threshold and the requirement that the second pulse has at least a $2 \times V_\mathrm{RMS}$ signal. This simulation includes all relevant effects such as a realistic neutrino vertex distribution, viewing angle differences of the two signal trajectories, incoming signal directions, reflection at the surface, etc. (see \cite{NuRadioMC2019} for details). 

Especially for neutrino energies of \SI{e17}{eV}, the detection efficiency decreases quickly with depth. Thus, the optimal depth for exploiting the D'n'R technique represents a compromise between energy resolution which increases with depth, and the fraction of neutrino events that have a D'n'R signature, which decreases with depth. As the energy resolution is ultimately limited by the unknown inelasticity, it is not required to be much better than this limit. Therefore, shallower depths are favored, for which the energy resolution is already well below the inelasticity limit, but the detection efficiency is still high. We estimate \SI{-15}{m} to be the optimal receiver depth.

\subsection{Systematic uncertainties}
This section briefly discusses the relevant sources of systematic uncertainties. The level of systematic uncertainty will depend on the exact experimental setup and the quality of calibration procedures. Hence, we will not be able to quantify systematic uncertainties but we will list the relevant parameters to guide the design of the station layout and calibration procedures of a future experiment. 

The depth of the receiver directly affects the translation from $\Delta t$ to vertex distance. The depth uncertainty should be small compared to the experimental uncertainty of $\Delta t$, i.e., $\sigma_d << \SI{0.2}{ns} \times c \approx \SI{46}{mm}$, where we used the speed-of-light for an index-of-refraction of $n = 1.3$. A change in the depth of the receiver due to snow accumulation can be monitored precisely using the setup described in Sec.~\ref{sec:snowaccumulation}.

Uncertainties in the index-of-refraction $n(z)$ profile propagate to a systematic uncertainty in the conversion from $\Delta t$ and zenith angle to distance. A good understanding of the ice properties and especially the change of the $n(z)$ profile in the upper part of the ice is mandatory. Fortunately, the D'n'R measurement itself in combination with calibration transmitters can be used to measure the $n(z)$ profile with adequate precision (cf. Sec.~\ref{sec:nz_measurement}).

The incoming signal direction is typically determined by the signal arrival times in multiple spatially separated antennas. Thus, the position of the antennas as well as the time delays from cables etc. are the relevant source of systematic uncertainty (cf.~\cite{GaswintICRC2019}). 

The direct and reflected signal path are, in general, launched at different angles with respect to the Cherenkov cone and thus the two signals might have a different frequency content which can complicate the experimental determination of the time delay between the two signals. Propagation effects (like frequency dependent attenuation) due to the additional path length of the reflected ray are negligible for typical vertex distances of beyond several hundreds of meters. 
The effect of a different frequency content is reduced by the relatively narrowband response of a dipole receiver. In general, the different launch angles and the correspondingly different frequency content of the two pulses might even be beneficial for the reconstruction as it adds sensitivity to the viewing angle (cf.~Eq.~\eqref{eq:y} and \cite{GlaserICRC2019}).

\section{Experimental test of D'n'R technique}
\label{sec:experiment}
We tested the feasibility to measure the direct and reflected pulse with the following in-situ measurement. We use one ARIANNA detector station installed at Moore's Bay on the Ross Ice shelf that is equipped with a dipole at a depth of \SI{-8.6}{m}. We drilled a \SI{20}{m} deep hole $\sim$\SI{40}{m} away from the station using a newly developed portable cylindrical electrothermal drill \citep{Heinen2017, MeltingProbe} - this technique allowed a fast setup and to drill the borehole for the dipole within a few hours and required very little monitoring.. We installed a dipole antenna at \SI{-18.2}{m} connected through a coax cable to an Avtech pulse generator. The emitting antenna is a copper fat-dipole of \SI{35}{cm} length and diameter of \SI{5}{cm}, providing ample clearance in the \SI{9}{cm} wide borehole. The receiving antenna of the ARIANNA station is a \SI{52}{cm} long dipole with a diameter of \SI{8}{cm} \cite{Kravchenko2007}. Deployment of the larger antenna here takes advantage of the larger-diameter hole that can be drilled at these relatively shallow depths.

\begin{figure}[t]
    \centering
    \includegraphics[width=0.7\textwidth]{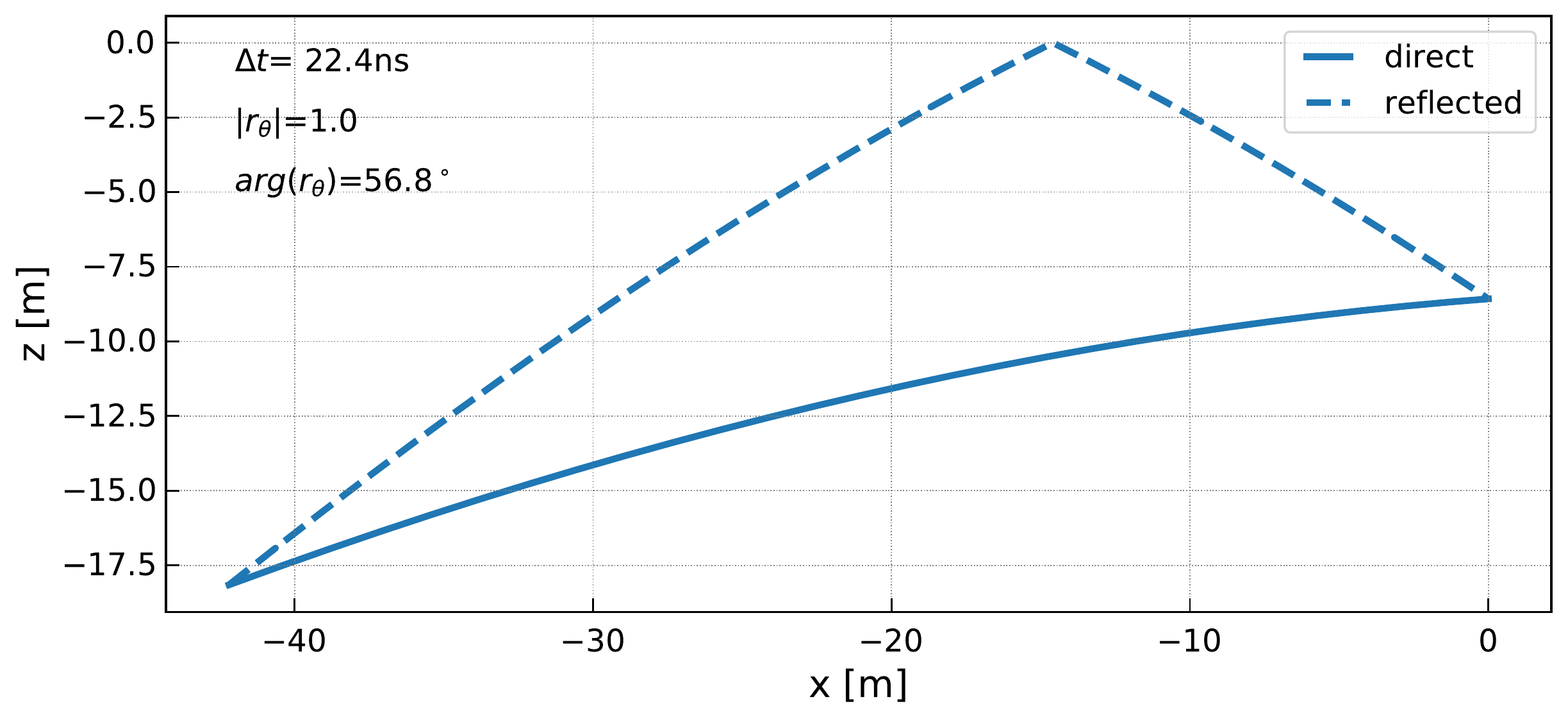}
    \caption{Signal paths from emitter to receiver. The legend shows the time difference between the reflected and direct signal path, as well as the magnitude and phase of the reflection coefficient for the $\theta$ polarization (p-wave).}
    \label{fig:path}
\end{figure}

The geometry of the measurement setup and the two signal paths from emitter to receiver are shown in Fig.~\ref{fig:path}. The signal paths are calculated with NuRadioMC \cite{NuRadioMC2019} using the \emph{Moore's Bay \#2} index-of-refraction profile from \cite{Barwick2018}.  The emitted signal is $\vec{e}_\theta$ polarized, as the emitting dipole is vertically oriented. We calculate the Fresnel reflection coefficient for the $\vec{e}_\theta$ polarization and find that the signal undergoes total internal reflection with a phase shift of \SI{57}{\degree}. The expected time delay between the reflected and direct signal is $\Delta t$ = \SI{22.4}{ns}. We also calculated the expected ratio of signal amplitudes taking into account the signal attenuation (which is proportional to the path length), and the directional sensitivity of the emitting and receiving dipole antennas. We predict an amplitude ratio between direct and reflected signal of $A_d/A_r = 1.66$.  We note that we expect equal D/R amplitudes for Askaryan signals from neutrinos as the emission point is much farther away than the depth of the receiver. 

\begin{figure}[t]
    \centering
    \includegraphics[width=0.49\textwidth]{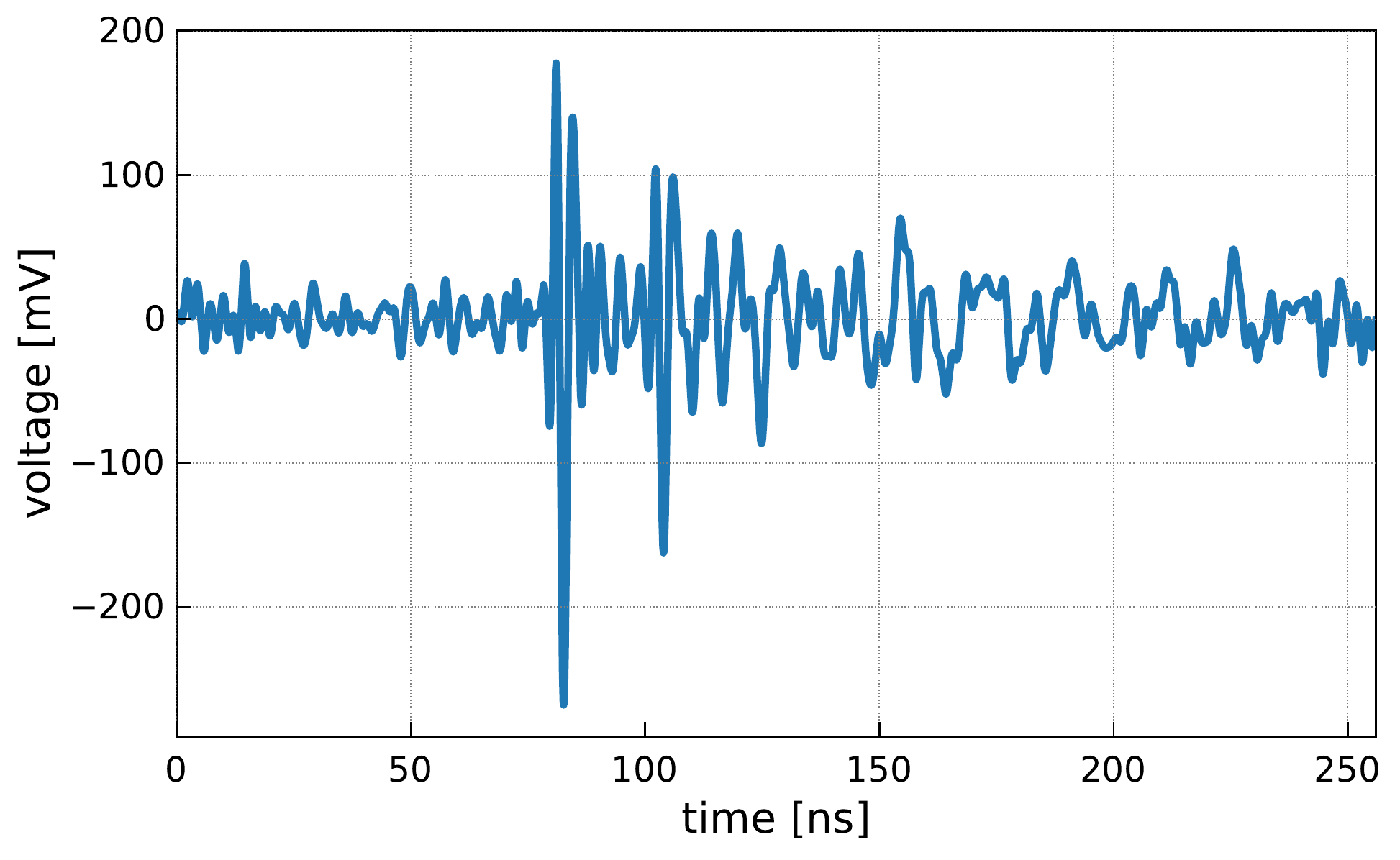}
    \includegraphics[width=0.49\textwidth]{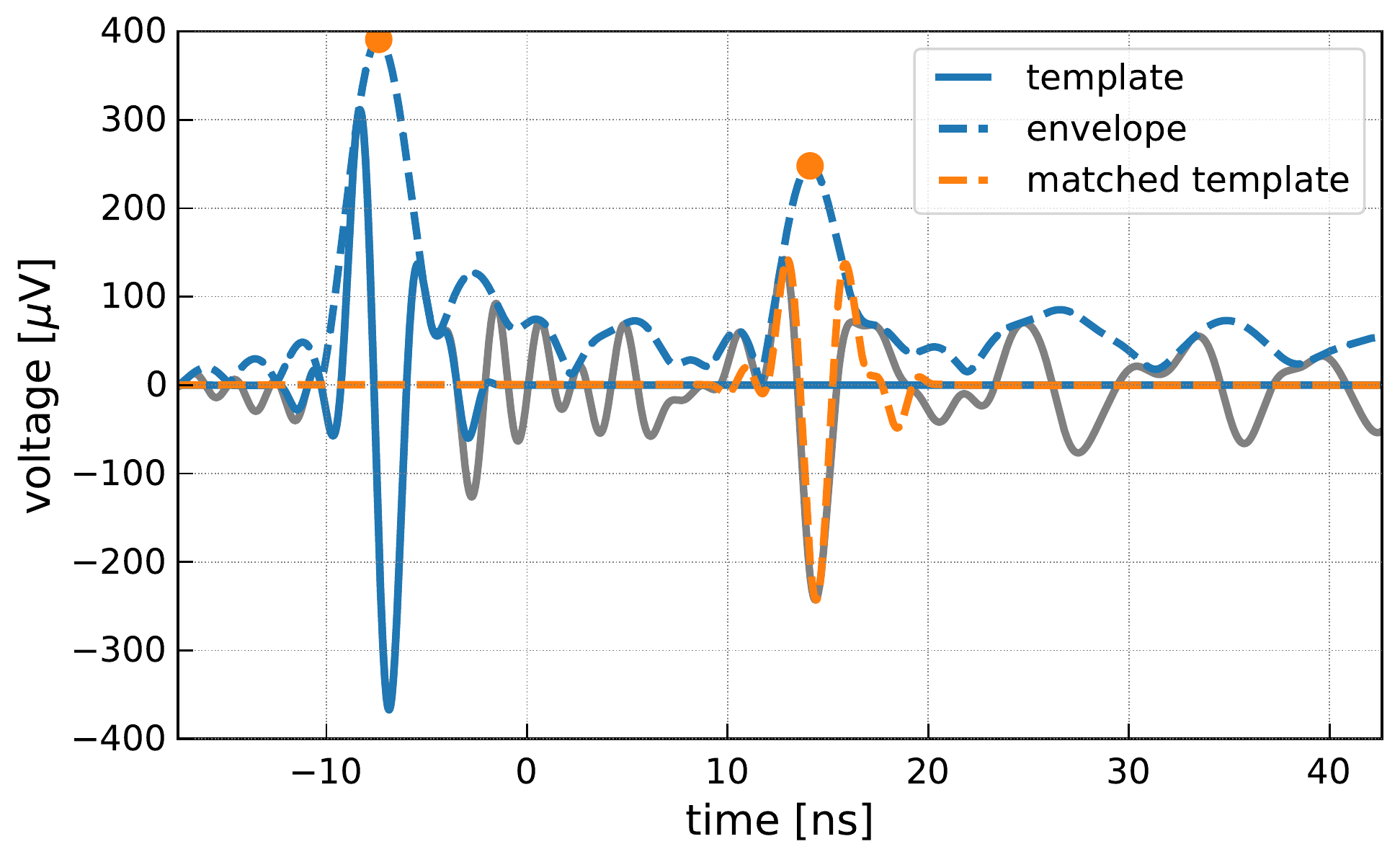}
    \caption{(left) Measured voltage trace of run 2 of the Avtech pulser run (see text for details). Shown is the voltage as a function of time. (right) Average over all run 2 events after correcting for the hardware response. The dashed blue line shows the Hilbert envelope of the voltage trace. The signal template is shown as the solid blue curve. The phase shifted template of the reflected pulse, for the best fit, is shown as the orange dashed curve.}
    \label{fig:trace_run2}
\end{figure}

We performed two measurement runs with different amplitudes. The first run was with a pulse amplitude of \SI{5}{V} and a repetition rate of \SI{0.5}{Hz}. The second run was with a pulse amplitude of \SI{2.5}{V} and a repetition rate of \SI{5}{Hz}. In both cases, the width of the pulse was set to \SI{0.5}{ns} FWHM. Data were taken for approximately \SI{10}{min} each, resulting in about 300 events for run 1, and 3500 events for run 2. The readout of the ARIANNA station was triggered externally by generating a trigger signal along with the pulse. Each recorded voltage trace has a physical length of \SI{256}{ns} sampled with \SI{1}{Gsample/s}. 
One of the recorded waveforms of run 2 is presented in Fig.~\ref{fig:trace_run2}. The direct and reflected pulses are clearly visible. 

We process the data using the NuRadioReco software \cite{NuRadioReco2019}, as follows. The voltage traces are upsampled at \SI{100}{Gsamples/s} and the amplifier and cable responses are unfolded. A 5th order Butterworth filter with a passband of \SI{100}{MHz} to \SI{450}{MHz} is applied to filter out noise outside of the signal bandwidth. The timing between the measurements is synchronized by shifting all traces in time to match the first recorded trace. Then, the average over all measured traces of each run is calculated. The result for run 2 is presented in the right panel of Fig.~\ref{fig:trace_run2}. 

The time delay $\Delta t$ is determined as follows: A template of the signal pulse is obtained by filtering out the first pulse using a modified Hanning window in the time domain. The Hanning window is adjusted to transition from 0 to 1 (and 1 to 0) within \SI{3}{ns} and set to $1$ for \SI{6}{ns}. The window is centered around the maximum of the Hilbert envelope (dotted blue curve and orange circle of Fig.~\ref{fig:trace_run2} (right)). The resulting signal pulse template is shown as the blue curve. Then, the signal template is adjusted for the phase shift of \SI{57}{\degree} resulting from the Fresnel reflection off the ice surface. The optimal time shift is found by determining the time shift value that maximizes the cross correlation between the matched template and the measured trace. The precision is further improved by fitting a Gaussian function to the cross correlation around the maximum. The matched template is shown as the orange dotted curve. We measure $\Delta t = \SI{21.743}{ns}$. The same procedure was repeated for run 1, yielding $\Delta t = \SI{21.762}{ns}$. Thus, the measurement is reproducible to within \SI{19}{ps} which is remarkable given that the data is sampled with only \SI{1}{Gsample/s}, corresponding to \SI{1}{ns} wide time bins. A comparison with theoretical expectation and a discussion of systematic uncertainties is presented later in Sec.~\ref{sec:avtech_sys}.

The amplitude ratio can be estimated either from the ratio of the maxima of the Hilbert envelope (cf. Fig.~\ref{fig:trace_run2} right), yielding $A_d/A_r$ = 1.43 for run 1 and $A_d/A_r = 1.58$ for run 2, or by scaling the amplitude of the reflected template to the minimum of the reflected pulse, yielding $A_d/A_r$ = 1.47 for run 1 and $A_d/A_r = 1.61$ for run 2. The determination of the amplitude ratio is more challenging because the reflected pulse interferes with after-pulses from the first direct pulse, altering the signal amplitude. The after-pulse hypothesis is supported by the observation of oscillations above the noise level after the first direct signal pulse, and the observation that the last part of the reflected template does not exactly follow the measurement. 
The measurements of the two runs agree within $\sim$10\%. Given this level of experimental uncertainty, the measurement is in agreement with the theoretical expectation of $A_d/A_r = 1.66$. We conclude that the ice surface acts as a flat reflector and specularly reflects the radio pulse without significant attenuation.

\subsection{Systematic uncertainties}
\label{sec:avtech_sys}
As illustrated above, one of the advantages of the D'n'R technique is that the time difference can be measured experimentally with high accuracy. The time difference is determined from a single voltage trace. Time synchronization between different channels, cable delays, amplifier characteristics, etc are irrelevant. Thus, systematic uncertainties of the experimental determination of $\Delta t$ are negligible.

However, several uncertainties influence the theoretical calculation of the expected $\Delta t$: Uncertainties in the geometry (the position of the emitter and receiver) influence the calculation; the modelling of the index-of-refraction profile also impacts the calculated signal paths and propagation times. 

The experimental uncertainties in the positioning of the emitting dipole antenna are \SI{20}{cm} in depth and \SI{10}{cm} in $x$ and $y$. The position of the receiving antenna was measured more precisely to \SI{10}{cm} accuracy in depth, and \SI{5}{cm} in $x$ and $y$. A variation in the depth of \SI{10}{cm} leads to a change in the predicted $\Delta t$ of \SI{0.14}{ns}, and a change of \SI{10}{cm} in the horizontal distance leads to a change in $\Delta t$ of \SI{0.08}{ns}. Changes in $\Delta t$ scale roughly linearly with the displacement. Thus, changing the position of receiver and emitter by one standard deviation can lead up to variations of  $\pm$\SI{0.56}{ns}.

\subsection{Measurement of density profile}
\label{sec:nz_measurement}
The index-of-refraction profile, which is directly linked to the density of the snow/ice via the Schytt equation, is described by an exponential function with two free parameters. These parameters were optimized to match in-situ index-of-refraction measurements of different depths. The exponential profile describes the data well but the parameters carry an uncertainty \cite{Barwick2018}. Varying the parameters within one standard deviation of their uncertainty leads to a maximum variation in the predicted $\Delta t$ of $\pm$\SI{0.38}{ns}. 

Hence, the difference between the predicted and measured $\Delta t$ of \SI{0.66}{ns} is compatible within the systematic uncertainties. In the future, with a more precise measurement of the geometry, this measurement can be used to precisely determine the index-of-refraction profile $n(z)$ near the surface. The $n(z)$ profile can be described with an exponential function
\begin{equation}
    n(z) = 1.78 - \Delta n \, e^{-z/z_0} \, 
\end{equation}
with the two free parameters $\Delta n$ and $z_0$. The parameters for the Moore's Bay site that were determined via an optimization to in-situ measurements are $\Delta n$ = \num{0.481\pm0.007} and $z_0$ = \SI{37\pm1}{m} \cite{Barwick2018}.
Thus, two independent $\Delta t$ measurements are required to determine both parameters. This can be achieved, e.g., with two receivers at \SI{10}{m} and \SI{20}{m} depth and an emitter \SI{40}{m} away and at a depth of \SI{15}{m}, similar to our measurement setup but with two receivers at different depths. 

We studied the achievable uncertainty in a toy MC assuming a $\Delta t$ resolution of \SI{10}{ps}. From the two $\Delta t$ measurements we can determine the parameter $\Delta n$ to a precision of $0.0006$ and $z_0$ to a precision of \SI{0.2}{m} which constitutes an order of magnitude improvement over the parameters derived from snow density measurements of \cite{Barwick2018}.

The resolution can be further improved by an improved geometry: The more distinct the ray paths to the two receivers, the better the sensitivity to the $n(z)$ parameters. Possible geometries are constrained by the demand of having two ray paths to the receiver and having total-internal-reflection at the surface. Increasing the distance between emitter and the shallower receiver (\SI{-10}{m}) to \SI{80}{m} improves the precision by another factor of $2$. In general, adding more receivers at different depths and distances would allow to determine the parameters of more elaborate $n(z)$ models, having more free parameters. 
Another prospect of multi receiver measurements is a long term measurement of the snow accumulation in conjunction with the $n(z)$ profile to track possible snow compaction at the surface. 

\subsection{Time resolution at low signal-to-noise ratios}
So far we have considered only measurements at relatively high signal-to-noise ratios and even an average over several events to improve the $\Delta t$ resolution. In the case of an neutrino signal, only one measurement is possible, at often small signal-to-noise ratios (SNRs) just above the trigger threshold. We now consider the evolution of the $\Delta t$ resolution with SNR. 

\begin{figure}[t]
    \centering
    \includegraphics[width=0.6\textwidth]{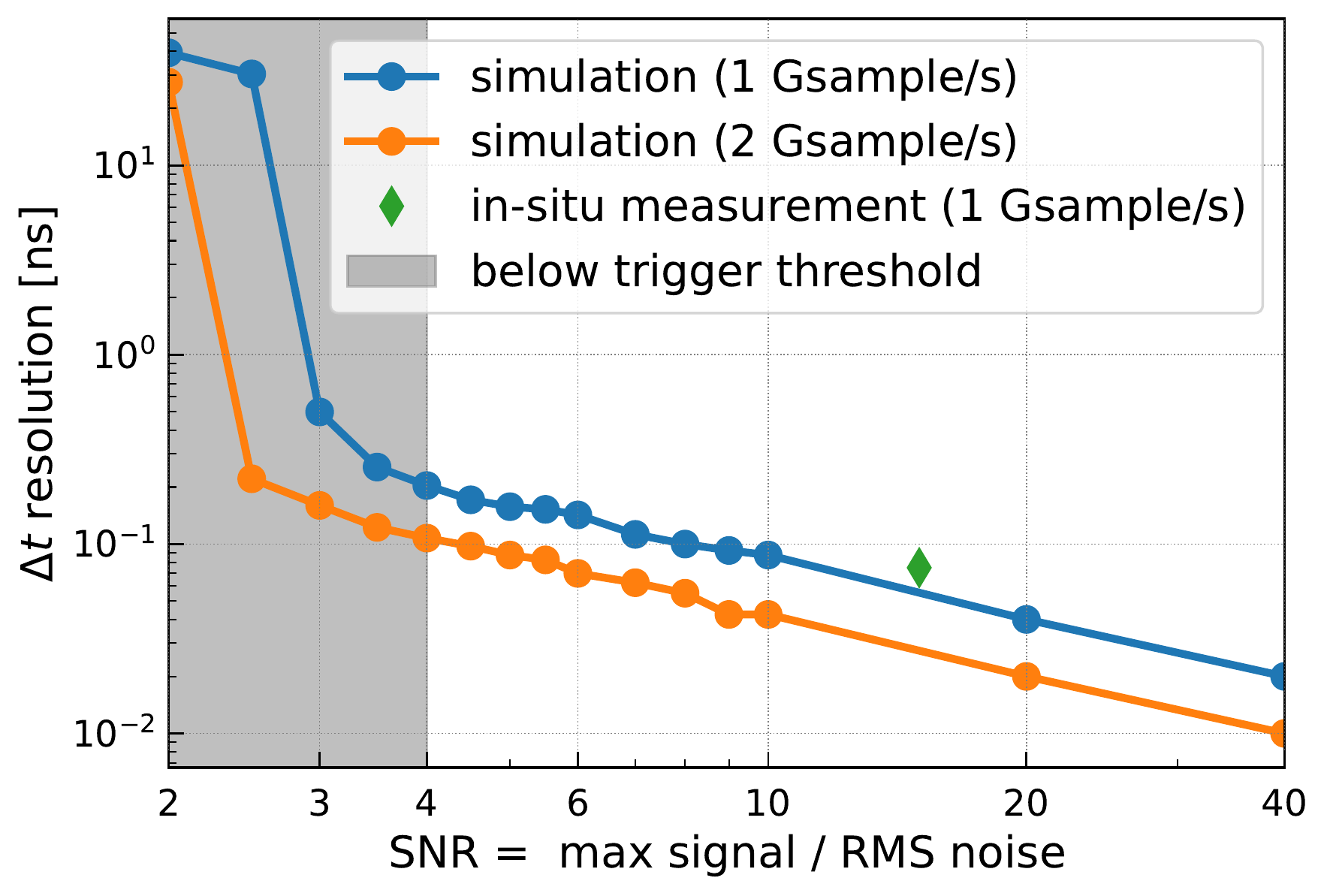}
    \caption{Resolution of time delay between direct and reflected pulse as a function of signal-to-noise ratio. Two simulations of different sampling frequencies are compared to an in-situ measurement.}
    \label{fig:DtSNR}
\end{figure}

The typical distance to the neutrino interaction vertex is typically far away from the detector (see Fig.~\ref{fig:MCdist} left and \cite{NuRadioMC2019}) such that both the direct and reflected pulse have a similar amplitude when received with a dipole antenna with equal sensitivity to updward and downward coming signals
equally displaced from the horizon. We calculate the predicted Askaryan signal using the precise \emph{ARZ2019} time-domain model \cite{NuRadioMC2019, Alvarez-Muniz2011}, fold this pulse with the antenna response, place two copies of the pulse \SI{10}{ns} apart, add thermal noise and calculate the time delay between the two pulses using the same cross-correlation method as discussed above. We repeat this a thousand times for each signal-to-noise ratio to estimate the $\Delta t$ resolution from the standard deviation of the time-delay distribution. The result is presented in Fig.~\ref{fig:DtSNR} for two different sampling frequencies of $1$ and \SI{2}{Gsamples/s}.

In addition, we calculate the $\Delta t$ resolution from the in-situ measurement. For each of the 3500 events of run 2 we calculate the time delay individually and estimate the uncertainty via the standard deviation of the $\Delta t$ distribution. The signal-to-noise ratio of the smaller second pulse is SNR=15; this measurement is also shown in Fig.~\ref{fig:DtSNR}. It is close to the simulated resolution but slightly higher indicating that noise is not the only source of uncertainty at these high SNRs. 

At typical trigger thresholds above $4 \times V_\mathrm{RMS}$ \cite{ARIANNA2019} and a sampling rate of \SI{2}{Gsamples/s}, the maximum $\Delta t$ uncertainty is \SI{0.1}{ns} and therefore well below the $\Delta t$ uncertainty of \SI{0.2}{ns} assumed in the simulation study of Sec.~\ref{sec:theory}. There are efforts to lower the trigger thresholds substantially to about $2 \times V_\mathrm{RMS}$ using a phased array \cite{Allison_2019}: Four to eight antennas are phased up to increase the signal-to-noise ratio by a factor of $\sqrt{\text{number of antennas}}$. Although this lowers the trigger threshold, the available signal-to-noise ratio for reconstruction is still above 4 as multiple antennas can be combined. Actually, a phased-array component at around \SI{15}{m} depth will be ideally suited to precisely measure the D'n'R time delay, and thus, provide an excellent vertex distance resolution.  

\section{Monitoring of snow accumulation}
\label{sec:snowaccumulation}

The time difference between the direct and reflected signal path is sensitive to the depth of the receiving antenna. Due to variations in snow accumulation, this depth changes over time. Thus, a precise monitoring of snow accumulation is crucial for the success of the D'n'R technique. Fortunately, the calibration measurement outlined in the previous section can be easily turned into a precise snow accumulation monitoring device. 

\subsection{Snow accumulation at the ARIANNA site}

The ARIANNA electronics are capable of creating a pulse similar to the capabilities of the Avtech pulse generator, with an FWHM pulse width of \SI{2}{ns}. 
This so-called heartbeat pulser was connected through two \SI{25}{m} LMR-600 and one \SI{20}{m} LMR-240 cable to the transmitting antenna. 
An additional \SI{100}{MHz} high pass filter was installed to prevent leakage of low-frequency noise from the electronics box.
Both the transmitting and receiving antenna are held in place by a rope with a small expansion coefficient that was attached to a bamboo pole which was stuck into the snow. 

The heartbeat pulser can be activated remotely. The data acquisition system was set up to start a heartbeat run every \SI{12}{hours}. For \SI{5}{min}, the heartbeat pulser was activated with a repetition rate of \SI{0.5}{Hz}, yielding 150 events per run. This periodic calibration run reduces the time in neutrino observation mode by less then 1\% and allows continuous monitoring of the snow accumulation. 

\begin{figure}[t]
    \centering
    \includegraphics[width=\textwidth]{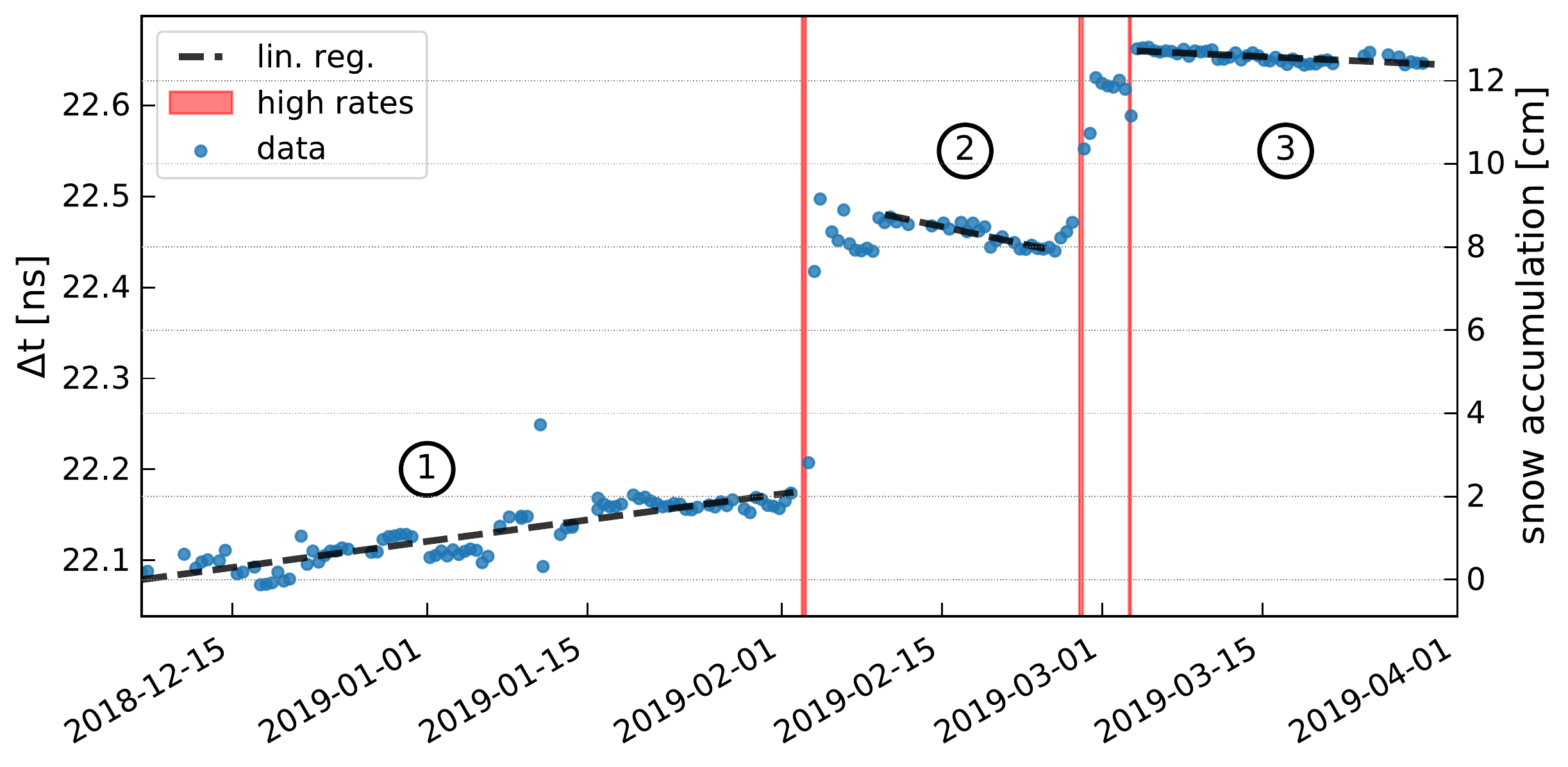}
    \caption{Monitoring of snow accumulation. Measured $\Delta t$ (left axis) is overlaid with the corresponding snow accumulation (blue circles, corresponding to the right axis) as a function of time. The red shaded areas show periods of extended high rates, an indicator of storms. The dashed black lines show a linear regression to the data points.}
    \label{fig:heartbeat}
\end{figure}

This calibration system ran from the beginning of December 2018 (after the deployment team left the site) until April 2019 when the station turned off as the solar panel output fell below threshold. We process the data in the same way as described in the previous section, but without the bandpass filter. Because of the larger pulse width, additional signal dispersion from the high pass filter and larger attenuation of high frequencies in the long coax cables, the heartbeat pulse has more low frequency content. However, this does not impact the capability to measure the $\Delta t$, as demonstrated below.
We average over each set of 150 events and calculate $\Delta t$ from the average trace. The measured time delays as a function of time are presented in Fig.~\ref{fig:heartbeat}.

\subsubsection{Interpretation}
The change in the measured $\Delta t's$ is converted to a change in the snow accumulation and shown on the right y-axis of Fig.~\ref{fig:heartbeat}. We observe three clear jumps in snow accumulation -- one at the beginning of February and two at the beginning of March. These times are correlated with an extended period of atypically high trigger rates in all ARIANNA stations.  This feature is known to correlate with high winds associated with storms. These observations support the interpretation that the change in $\Delta t$ is due to an increase in the snow level.

The Fresnel reflection zone has a radius of several meters. Thus, our measurement probes the average snow accumulation over an extended area around the point of reflection determined by the ray-tracing model (cf. Fig.~\ref{fig:path}). The present measurement does not distinguish between precipitation or snowdrift.  

The snow surface at Moore's Bay is not perfectly flat but exhibits variations in height of a few centimeters over meter scales.  These features are present near the ARIANNA station, but larger amplitude features, such as tilts or Sastrugi, are absent. Possible timing errors introduced by the averaging of small scale surface features within the Fresnel reflection zone are anticipated to be insignificant. 

For neutrino studies, the position of the snow surface must be known for any relevant reflection point.  Though we expect that the observed patch of snow surface is representative of any nearby surface position, this hypothesis must be confirmed by additional study. For example,  azimuthal symmetry can be tested by installing several emitters at opposite sides of the station. 

We subdivide the data into three time ranges where no large snow accumulation occurred and describe the average trend with a straight line obtained via a linear regression. In the first period, the snow accumulation slowly and constantly rose by \SI{2.6}{mm/week}. A plausible explanation for this effect is that piles of snow left by the deployment teams are re-distributed due to wind giving rise to a slight increase of the snow overburden. In the second and third time periods, we observe a reduction in snow accumulation by \SI{-4.1}{mm/week} and \SI{-0.9}{mm/week}. This is likely because fresh snow compacts over time due to solar illumination. This also explains why the slope in the third period is smaller: At this time of year the number of sunlight hours per day is decreasing and the sun is at a lower declination. We note that a change in the density of the new snow layer corresponds to a change in the index-of-refraction which will effect the signal trajectories and thereby the relation between $\Delta t$ and snow accumulation. However, this effect should be small compared to the reduction in snow accumulation itself. A setup with multiple receivers at different positions would allow to disentangle these effects (cf. discussion in Sec.~\ref{sec:nz_measurement}) and is foreseen for the future.

We estimate the resolution in this method by calculating the scatter around the linear trend in the three time periods. We measure a scatter of \SI{18}{ps} in the first period and scatters of \SI{5.6}{ps} and \SI{4.1}{ps} in the second and third period, respectively. This translates into a resolution in the snow accumulation of \SI{4}{mm} and $\sim$\SI{1}{mm}. This resolution is an order of magnitude lower than the assumed $\Delta t$ resolution of Sec.~\ref{sec:theory} of \SI{0.2}{ns}. Thus, changes in snow accumulation do not present any limitation on  the D'n'R technique, can be corrected for, and do not deteriorate the energy resolution.

\subsection{Synergies with geophysics}
A construction of a large scale radio neutrino detector with $\mathcal{O}$(100) stations separated by \SI{1.5}{km} is foreseen for the future (see e.g. \cite{ARIANNACospar2018}). If each station is equipped with the functionality to measure the snow accumulation, a continuous large scale monitoring of the surface mass balance is possible which is crucial goal in geophysics and important for predicting future behavior of ice sheets and their effect on sea level \cite{Lenaerts2019}. With a precision of \SI{1}{mm} and measurement timescales measured in hours, our measurement is competitive with, or possibly even more precise than, traditional methods. A high-energy neutrino detector can thus also contributed to geophysics and the monitoring of ice sheets in Arctic regions.

\section{Discussion and Conclusions}
The D'n'R technique provides unique opportunities for the detection of high-energy neutrinos. An antenna placed $\mathcal{O}$(\SI{15}{m}) below the ice surface will measure a direct pulse and a pulse reflected off the ice surface for most detected neutrino interactions in the ice. This signature does not only provide a unique characteristic of a neutrino origin of the signal but also allows to measure the distance to the neutrino interaction vertex precisely, a crucial property to determine the neutrino energy.

With reasonable assumptions on the uncertainties of the zenith angle of \SI{0.2}{\degree} and time delay between the two pulses of \SI{0.2}{ns}, we find a vertex distance resolution of better than 12\%. The good timing resolution is possible because the time delay of two pulses within the same channel is calculated. Thus, most experimental uncertainties normally affecting the timing (such as cables delays, antenna differences, position of receivers, time synchronization of channels etc.) cancel out. In addition, the time delay and the zenith angle can be reconstructed independently and are thus uncorrelated. These two properties are the main advantage compared to a reconstruction of the shower front from the pulse arrival times of spatially displaced antennas where the distance, zenith and azimuth angles need to be determined simultaneously and are correlated with each other. 

The neutrino energy depends on the measured quantities energy fluence, distance, viewing angle and inelasticity. The stochastic process of the latter imposes a irreducible energy uncertainty of a factor of two. The vertex distance uncertainty of the D'n'R technique translates into a contribution to the energy uncertainty of 20\% (at $E_\nu = \SI{e17}{eV}$) and $\sim$40\% (at $E_\nu = \SI{e18}{eV}$). Thus, it is much smaller than the inelasticity limit and does not increase the energy uncertainty significantly. Also the uncertainties of the remaining contributions to the neutrino energy are smaller than the inelasticity limit as estimated in \cite{GlaserICRC2019}. Hence, a shallow high-energy neutrino detector will have an excellent energy resolution limited by the irreducible inelasticity fluctuations. 

The D'n'R technique was tested experimentally at Moore's Bay on the Ross ice shelf. An ARIANNA detector station equipped with a \SI{8.6}{m} deep dipole antenna received signals from a \SI{20}{m} deep transmitter. The observed time delay, the amplitude ratio as well as the phase shift of the reflected pulse matches the expectation within systematic uncertainties providing an important proof-of-concept of the method. We found a time resolution of \SI{80}{ps} for the signal-to-noise ratio of 15 of the measurement. The uncertainty decreases slightly with smaller signal-to-noise ratios but remains below \SI{0.2}{ns} which was determined in a simulation study. 

This test setup was also used as a snow accumulation monitoring device: the ARIANNA hardware is equipped with a remotely configurable pulse generator that was connected to the transmitting antenna. The D'n'R time delay was measured every \SI{12}{h} over several months which was translated into snow accumulation, as the time delay increases proportional to the snow accumulation. From this long-term measurement, we derived a statistical uncertainty of \SI{5}{ps} corresponding to a \SI{1}{mm} resolution on the snow accumulation. A large scale neutrino detector equipped with such a calibration device might contribute important data on the surface mass balance of ice sheets, a crucial property for global climate models.

\section{Acknowledgements}
We are grateful to the U.S. National Science Foundation-Office of Polar Programs, the U.S. National Science Foundation-Physics Division (grant NSF-1607719) for granting the ARIANNA array at Moore's Bay. Without the invaluable support of the people at McMurdo, the ARIANNA stations would have never been built.

We acknowledge funding from the German research foundation (DFG) under grants GL 914/1-1 (CG), NE 2031/1-1, and NE 2031/2-1 (DGF, ANe, IP, CW) and the Taiwan Ministry of Science and Technology (JN, SHW).  HB acknowledges support from the Swedish Government strategic program Stand Up for Energy.  EU acknowledges support from the Uppsala university Vice-Chancellor's  travel  grant  (sponsored  by  the  Knut  and  Alice  Wallenberg  Foundation)  and  the  C.F. Liljewalch travel scholarships.  DB and ANo acknowledge support from the MEPhI Academic Excellence Project (Contract No.  02.a03.21.0005) and the Megagrant 2013 program of Russia, via agreement 14.12.31.0006 from 24.06.2013.

The used melting probes (cf. Sec.~\ref{sec:experiment}) were developed within the EnEx-RANGE project, which was funded by the German Federal Ministry of Economics and Energy (BMWi) by resolution of the German Federal Parliament under the funding code 50NA1501.

\bibliographystyle{JHEP}
\bibliography{BIB}

\appendix
\end{document}